\documentclass{sig-alternate}

\usepackage{url}
\usepackage{amsmath}
\usepackage{amssymb}
\usepackage{bm}
\usepackage{stmaryrd}
\usepackage{color}
\usepackage{colortbl}
\usepackage{epsf}
\usepackage{supertabular}

\usepackage{multicol}

\newcommand{\ignore}[1]{}

\ignore{
\newtheorem{proposition}{Proposition}
\newtheorem{lemma}{Lemma}

\newtheorem{theorem}{Theorem}
\newtheorem{definition}{Definition}

\newtheorem{example}{Example}
}

\DeclareMathAlphabet{\mathpzc}{OT1}{pzc}{m}{it}

\def\be{\begin{equation}}
\def\ee{\end{equation}}
\def\bea{\begin{eqnarray}}
\def\eea{\end{eqnarray}}
\def\nn{\nonumber}

\def\s{\sigma}

\def\l{\left}
\def\r{\right}

\def\w{\wedge}
\def\ra{\rightarrow}

\def\P{\Phi}

\def\fa{\forall}

\def\bc{\bigcap}
\def\S{\Sigma}
\def\nin{\not\in}

\def\ex{\exists}

\def\hb{\hbox}
\def\lar{\leftarrow}

\def\sc{\scriptsize}

\def\ba{\begin{align}}
\def\ea{\end{align}}
\def\bes{\begin{split}}
\def\es{\end{split}}
\def\vD{\vDash}

\def\ems{\emptyset}

\newcommand{\red}[1]{{\em \bf\textcolor[rgb]{1.00,0.00,0.00}{#1}}}

\newcommand{\comlb}[1]{{\vspace{2mm}\noindent \bf \red{COMM(LEO):}}~ #1 \hfill {\bf
    END.}\\}
\newcommand{\comj}[1]{{\vspace{2mm}\noindent \bf \red{COMM(JAF):}}~ #1 \hfill {\bf
    END.}\\}

\newcommand{\defproof}[2]{{\noindent\bf Proof of #1:\
}#2 \boxtheorem}

\newcommand{\boxtheorem}{\hfill $\Box$}
\newcommand{\nit}[1]{{\it #1}}

\newcommand{\mdg}{\nit{MDG}(M)}
\newcommand{\mc}[1]{\mathcal{ #1}}

\begin{document}




\title{Query Answering under
Matching Dependencies for Data Cleaning: Complexity and Algorithms}
\numberofauthors{2}
\author{
\alignauthor Jaffer Gardezi\\
\affaddr{University of Ottawa, SITE}\\
\affaddr{Ottawa, Canada}\\
\email{{\small \bf jgard082@uottawa.ca}}
\alignauthor Leopoldo Bertossi\\
\affaddr{Carleton University, SCS}\\
\affaddr{Ottawa, Canada}\\
\email{\small \bf bertossi@scs.carleton.ca}
}


\maketitle

\begin{abstract}
Matching dependencies (MDs) have been recently introduced as declarative rules for entity
resolution (ER), i.e. for identifying and resolving duplicates in relational instance $D$.
A set of MDs can be used as the basis for a possibly non-deterministic mechanism that
computes a duplicate-free instance from $D$. The possible results of this process are the clean,
{\em minimally resolved instances} (MRIs). There might be several MRIs for $D$, and the
{\em resolved answers} to a query are those that are shared by all the MRIs. We investigate the
problem of computing resolved answers.
We look at various sets of
MDs, developing syntactic criteria for determining
(in)tractability of the resolved answer problem, including a dichotomy result. For some tractable
classes  of MDs and conjunctive queries, we
present a query rewriting methodology that can be used to
retrieve the resolved answers.
We also investigate connections with {\em consistent query answering},
deriving further tractability results
for MD-based ER.
\end{abstract}




\ignore{\newtheorem{theorem}{Theorem}
\newtheorem{lemma}{Lemma}
\newtheorem{corollary}{Corollary}
\newtheorem{proposition}{Proposition}}
\newdef{theorem}{Theorem}
\newdef{lemma}{Lemma}
\newdef{corollary}{Corollary}
\newdef{proposition}{Proposition}
\newdef{definition}{Definition}
\newdef{example}{Example}

\section{Introduction}

For different reasons,  databases
may contain different coexisting representations of the same external, real world entity. Those duplicates can be entire tuples or values within them. Ideally, those tuples or values should be
merged into a single representation. Identifying and merging
duplicates is  a process called {\em entity resolution} (ER) \cite{naumannACMCS,elmargamid}. Matching
dependencies (MDs) are a recent proposal for declarative duplicate resolution \cite{Fan08,Fan09}.
An MD expresses,  in the form of a rule,
that if the values of certain attributes in a pair of
tuples are similar, then the values of other attributes in those tuples should be
matched (or merged) into a common value.

For example, the MD $R_1[X_1] \approx R_2[X_2]  \rightarrow  R_1[Y_1] \doteq R_2[Y_2]$
says that if an $R_1$-tuple and $R_2$-tuple have similar values for attributes $X_1,X_2$, then their values for $Y_1, Y_2$
should be made equal. This is a {\em dynamic} dependency, in the sense that its satisfaction is checked against a pair
of instances: the first where the antecedent holds and the second where the identification of values takes place.
This semantics of MDs was sketched in \cite{Fan09}.

The original semantics was refined
in  \cite{icdt11}, including the use of {\em matching functions} to do
the matching of two attribute values. Furthermore, the minimality of changes (due to the matchings) is guaranteed by means of
a chase-like procedure that changes values only when strictly needed.

An alternative refinement
of the original  semantics was proposed in  \cite{jaf11}, which is the basis for this work. In this case,
arbitrary values can be used for the matching. The semantics is also based on a chase-like procedure.
However, the minimality of the number of changes is explicitly imposed. In more detail,
in order to obtain a clean instance, an iterative procedure
is applied, in which the MDs are applied repeatedly. At each step,
merging of duplicates can generate additional similarities
between values, which forces the MDs to be applied again and
again, until a clean instance is reached.
Although MDs indicate values to be merged, the clean instance
obtained by applying this iterative process to a dirty instance will in general
depend on how the merging is done, and MDs do not specify this.
As expected, MDs can be applied in different orders. As a consequence, alternative clean
instances can be obtained. They are defined in \cite{jaf11} as the {\em minimally resolved instances} (MRIs).

Since there might be large portions of data that are not affected by
the occurrence of duplicates or by the entity resolution process, no matter how it is applied,
it becomes relevant to characterize and obtain those pieces of data that are invariant under the cleaning
process. They could be, in particular, answers to queries.  The {\em
resolved answers} \cite{jaf11} to a query posed to the original, dirty database are those answers to the query
that are invariant under the entity resolution process. In
principle, the resolved answers could be obtained by computing all the MRIs, and posing the query to all
of them, identifying later the shared answers. This may be too costly, and more efficient alternatives should be used
whenever possible, e.g. a mechanism that uses only the original, dirty instance.

In \cite{jaf11}, the problem of computing
resolved answers to a query was introduced, and
some preliminary and isolated complexity results were given.
In this work we largely extend those results on resolved query answering,
providing new complexity results, in Sections \ref{subsec:com} and \ref{subsec:NI2}.
For tractable cases, and for the  first time, a
query rewriting methodology for
efficiently retrieving the resolved answers is presented, in Section \ref{sec:Tr}.

Summarizing, in this paper, we undertake the first systematic investigation of the complexity of the problems of computing and deciding
resolved answers to conjunctive queries.  More, precisely, the contributions of this paper are
as follows:
\begin{enumerate}

\item Starting with the simplest cases of MDs and queries, we consider
 the complexity of computing the resolved
answers. We provide syntactic characterizations of easy and hard cases of MDs.

\item For certain sets of two MDs, we establish a dichotomy result, proving that deciding the
resolved answers is
in {\it PTIME} or
${\it NP}$-hard in data.

\item We then move on to larger sets of MDs,
establishing, in particular, tractability for some interesting cyclic sets of
MDs.

\item We consider the problem of retrieving the resolved
answers to a query by querying the original dirty database instance.
For tractable classes of MDs, and a class of first-order conjunctive queries, we show
that a query can be rewritten into a new query that,
posed to the original dirty instance,
returns the resolved answers to the original query. Although the rewritten query is not
necessarily first-order, it can be expressed in positive Datalog with recursion and counting,
which can be evaluated in polynomial time.

\item We establish a connection between MRIs and {\em database repairs} under key constraints
 as found in {\em consistent query answering} (CQA) \cite{Arenas99,B2006,chom07}. In CQA, the
repair semantics is usually based on deletion of
whole tuples, and minimality on comparison under set inclusion. Reductions
from/to CQA allow us to profit from results
for CQA, obtaining additional (in)tractability results for resolved query computation under MDs.

These intractability results are important in that
they show that our query rewriting methodology in 4. does not
apply to all conjunctive queries. On the other hand, the tractable
cases identified via CQA differ from those in item 4.: The class
of MDs is more restrictive, but the class of
conjunctive queries is larger.
\end{enumerate}
Our complexity analysis sheds some initial light on the intrinsic computational limitations of retrieving
the information from a dirty database that is invariant under entity resolution processes, as captured by
MDs.

The structure of the paper is as follows. Section \ref{sec:pre} introduces
notation used in the paper and reviews necessary material
from previous publications. Section \ref{subsec:com} investigates the complexity
of the problem of computing resolved answers, identifying various tractable and intractable
cases. In Section \ref{sec:Tr}, an efficient query rewriting methodology for obtaining
the resolved answers (in tractable cases) is described.
Section \ref{subsec:NI2} establishes the connection with CQA.
In Section \ref{sec:con} we draw some final conclusions.


\section{Preliminaries}
\label{sec:pre}

We consider a relational schema $\mathcal{S}$ that includes an enumerable, possibly infinite domain $U$, and a set $\mc{R}$ of database predicates. $\mathcal{S}$ determines a first-order (FO) language $L(\mathcal{S})$. An instance
$D$ of $\mathcal{S}$ is a finite set of ground atoms of the form $R(\bar{t})$, with $R \in \mc{R}$, say of arity $n$, and $\bar{t} \in U^n$. $R(D)$ denotes the extension
of $R$ in $D$. The set of all attributes of $R$ is
denoted by $\nit{attr}(R)$. We sometimes refer to attribute $A$ of  $R$ by $R[A]$.
We assume that all the attributes are different, and that we can identify attributes with {\em positions} in predicates,
e.g. $R[i]$, with $1 \leq i \leq n$.
If the $i$th attribute of predicate $R$ is $A$, for a tuple $t = (c_1, \ldots, c_n) \in R(D)$,
$t_R^D[A]$ (usually, simply $t_R[A]$ or $t[A]$ if the instance is understood) denotes the value $c_i$. The symbol $t[\bar A]$ denotes the tuple whose entries are the
values of the attributes in $\bar A$. Attributes have and may share subdomains that are contained
in $U$. 

In order to compare instances, obtained from the same instance through changes of attribute values, we use {\em tuple identifiers}: Each database tuple $R(c_1,\ldots,c_n) \in D$ has an identifier, say \nit{t},
making the tuple implicitly become $R(t,c_1,$ $\ldots,c_n)$. The $t$ value is taken by an additional
attribute, say $\nit{T}$, that acts as a key.  Identifiers are not subject to updates, and are
usually left
implicit. Sometimes we do not distinguish between a tuple and its tuple identifier. That is,
with now $t$ a tuple identifier (value), $t^D_R$ denotes the tuple $R(c_1,\ldots,c_n)$ above;
 and  $t^D_R[A_i]$, the value for attribute $A_i$, i.e. $c_i$ above.\footnote{If there there is not danger of confusion, we sometimes omit $D$ or $R$
from $t^D_R$, $t^D_R[A]$.} Two instances over the same schema that share the same tuple identifiers
are said to be {\em correlated}. In this case it is possible to unambiguously compare their tuples.

A {\em matching dependency} (MD) \cite{Fan08}, involving predicates\linebreak $R(A_1,\ldots,A_n)$, $S(B_1,\ldots,B_m)$, is
 a rule, $m$, of the form
{\small \bea \label{eq:md}
m\!: \ \ \bigwedge_{i \in I, j \in J}R[A_i] \approx_{ij} S[B_j] \ \ra \bigwedge_{i \in I', j \in J'}R[A_i]\doteq S[B_j].
\eea}
\hspace*{-2mm} The set of attributes on the left-hand-side (LHS) of $m$ (wrt the arrow) is denoted with $\nit{LHS}(m)$. Similarly for
the right-hand-side. The domain-dependent binary relations $\approx_{ij}$
denote similarity of attribute values from a shared domain. The symbol $\doteq$ means
that the values of the pair of attributes in $t_1$ and $t_2$
should be updated to the same value. In consequence, the intended semantics of the MD is that if
any pair of tuples, $t_1\in R(D)$ and $t_2\in S(D)$, satisfy the similarity conditions on the LHS, then for the same tuples
the attributes indicated on the RHS have to take the same values \cite{Fan09}.\footnote{We assume that
instances and MDs share the same schema.}
The similarity relations, generically denoted with $\approx$, are {\em symmetric}, and {\em reflexive}.
We assume  that all sets $M$ of MDs are in {\em standard form},
i.e. for no two different MDs $m_1, m_2 \in M$, $\nit{LHS}(m_1) = \nit{LHS}(m_2)$. All sets of MDs can
be put in this form.

For abbreviation, we will sometimes write MDs as
\bea
R[{\bar A}] \approx S[{\bar B}]\ra R[\bar C]\doteq S[\bar E], \label{eq:simplMD}
\eea
with $\bar A = (A_1, ...,A_k)$, $\bar B = (B_1, ...,B_k)$, $\bar C = (C_1, ..., C_{k'})$, and $\bar E = (E_1, ...,E_{k'})$ lists of attributes.
The pairs $(A_i,B_i)$ and $(C_i,E_i)$
are called  {\em corresponding pairs} of attributes in $({\bar A}, {\bar B})$ and $({\bar C}, {\bar E})$,
resp. For an instance $D$ and a pair of tuples $t_1\in R(D)$ and $t_2\in S(D)$,
$t_1[\bar A]\approx t_2[\bar B]$ indicates that the similarities of the values for all corresponding pairs
of attributes of $(\bar A,\bar B)$ hold. The notation $t_1[\bar C] \doteq t_2[\bar E]$ is used similarly.

\begin{definition}\cite{jaf11}
For  a set $M$ of MDs, the {\em MD-graph}, $\nit{MDG}(M)$, is a directed graph with a vertex
$m$ for each $m\in M$, and with an edge from
$m_1$ to $m_2$ iff \ $\nit{RHS}(m_1)\cap \nit{LHS}(m_2)\neq \emptyset$.
\boxtheorem
\end{definition}
MD-graphs can have self-loops. If the MD-graph
of a set of MDs contains edges it is called
{\em interacting}. Otherwise, it is called {\em non-interacting}.

Updates as prescribed by an MD are not  arbitrary. The allowed updates
are the matching of values when the preconditions are met, which is captured by the set of {\em modifiable
values}.
\begin{definition}\label{def:mod}
Let $D$ be an instance, $R \in \mathcal{R}$,
$t_R \in R(D)$, $C$ an attribute of $R$, and $M$ a set of MDs.
Value $t_R^D[C]$ is \textit{modifiable}
if there exist $S\in \mathcal{R}$,
$t_S\in S(D)$, an $m\in M$ of the form
$R[\bar A]\approx S[\bar B]\ra R[\bar C]\doteq S[\bar E]$,
and a corresponding pair $(C,E)$
of $(\bar C,\bar E)$, such that
one of the following holds: \ 1. $t_R[\bar A]\approx t_S[\bar B]$, but
$t_R[C]\neq t_S[E]$. \\ 2. $t_R[\bar A]\approx t_S[\bar B]$ and $t_S[E]$ is modifiable.
Value $t_R[C]$ is \emph{potentially modifiable} if
$t_R[\bar A]\approx t_S[\bar B]$ holds.
For a list of attributes $\bar C$, $t_R[\bar C]$ is (potentially) modifiable iff
there is a $C$ in $\bar C$ such that $t_R[C]$ is (potentially) modifiable.
\boxtheorem
\end{definition}

\begin{definition}\label{def:new} \cite{jaf11}
Let $D$, $D'$ be correlated instances, and $M$ a set of MDs.
$(D,D')$ satisfies $M$, denoted $(D,D')\vD M$, iff: \ \
1. For any pair of tuples $t_R\in R(D)$,
$t_S\in S(D)$, if there exists an $m  \in M$ of the form
$R[\bar A]\approx S[\bar B]\ra R[\bar C]\doteq S[\bar E]$ and
$t_R[\bar A]\approx t_S[\bar B]$, then for the
corresponding tuples $t_R'\in R(D')$ and $t_S'\in S(D')$,
it holds $t_R'[\bar C] = t_S'[\bar E]$.\\
2. For any tuple $t_R\in R(D)$
and any attribute $G$ of $R$,
if $t_R[G]$ is {\em non-modifiable}, then
$t_R'[G] = t_R[G]$. \boxtheorem
\end{definition}
This definition  of
MD satisfaction departs from \cite{Fan09}, which requires that updates preserve
similarities. Similarity preservation may force undesirable
changes \cite{jaf11}. The existence of the
updated instance $D'$ for $D$ is guaranteed \cite{jaf11}. Furthermore, wrt \cite{Fan09},
our definition does not
allow unnecessary changes from $D$ to $D'$.
Definitions \ref{def:mod} and \ref{def:new} require that only values of
attributes that appear on RHS of the arrow in some MD
are subject to updates. This motivates the following definition.
\begin{definition}\label{def:changeable}
For a set $M$ of MDs defined on
schema $\mc{S}$,
the {\em changeable
attributes} of $\mc{S}$ are those that appear to the right of the arrow
in some $m \in M$. The other attributes of $\mc{S}$
are called {\em unchangeable}.
\boxtheorem
\end{definition}
Definition \ref{def:new} allows us to define a
{\em clean instance} wrt $M$ as the result
of a sequence of updates, each step being satisfaction preserving,
leading to a {\em stable instance} \cite{Fan09}.

\begin{definition}\label{def:res} \cite{jaf11}
A {\em resolved instance} for
$D$ wrt $M$ is an instance $D'$,
such that there is sequence of instances
$D_1,D_2,...D_n$ with:
$(D,D_1)\vD M$, $(D_1,D_2)\vD M$,...,
$(D_{n-1},$ $D_n)\vD M$, $(D_n,D')\vD M$, and
$(D',D')\vD M$. ($D'$ is {\em stable}.)
\boxtheorem
\end{definition}

\begin{example} \label{ex:exp}
Consider the MD $R[A]\approx R[A]\ra R[B]\doteq R[B]$  on predicate $R$, and an instance
$D$: \vspace{1mm}
\begin{multicols}{2}
\begin{center}
\begin{tabular}{c|c|c|} \hline
$R(D)$&$A$ & $B$  \\ \hline
$t_1$&$a_1$ & $c_1$  \\
$t_2$&$a_1$ & $c_2$  \\
$t_3$&$b_1$ & $c_3$  \\
$t_4$&$b_1$ & $c_4$ \\ \cline{2-3}
\end{tabular}
\end{center}
\noindent It has several resolved instances, among them, four that minimize the number of changes. One of them is
$D_1$ below. A resolved  in-\linebreak
\end{multicols}
\vspace{-7mm}stance  that is not minimal in this sense is $D_2$.
\begin{center}
\begin{tabular}{c|c|c|}\hline
$R(D_1)$&$A$ & $B$  \\ \hline
$t_1$&$a_1$ & $c_1$  \\
$t_2$&$a_1$ & $c_1$  \\
$t_3$&$b_1$ & $c_3$  \\
$t_4$&$b_1$ & $c_3$ \\ \cline{2-3}
\end{tabular}
~~~~~~~~~~~~\begin{tabular}{c|c|c|}\hline
$R(D_2)$&$A$ & $B$  \\ \hline
$t_1$&$a_1$ & $c_1$  \\
$t_2$&$a_1$ & $c_1$  \\
$t_3$&$b_1$ & $c_1$  \\
$t_4$&$b_1$ & $c_1$ \\ \cline{2-3}
\end{tabular}
\end{center} \vspace{-5mm}\boxtheorem
\end{example}
As suggested by the previous example,  we will require that
the number of changes wrt instance $D$ are minimized.

\begin{definition}\label{def:changes}
For an instance $D$ of schema $\mc{S}$,\\
(a) $T_D := \{(t,A)~|~ t \mbox{ is the id of a tuple in } D \mbox{ and } A \mbox{ is an}$\\
$\mbox{attribute of the tuple}\}$.\\
(b) $f_D: T_D \rightarrow U$ is given by: $f_D(t,A) :=
\mbox{ the value for } A$\\
$\mbox{ in the tuple in } D \mbox{  with id } t$.\\
(c) For an instance $D'$ with the same tuple ids as $D$,\\
\hspace*{5mm}$S_{D,D'} := \{(t,A)\in T_D~|~ f_D(t,A) \neq f_{D'}(t,A)\}$. \boxtheorem
\end{definition}

\begin{definition} \label{def:minim} \cite{jaf11}
 A {\em minimally resolved instance} (MRI) of $D$
wrt $M$ is a resolved instance
$D'$ such that $|S_{D,D'}|$ is minimum, i.e.
there is no resolved
instance $D''$ with $|S_{D,D''}| < |S_{D,D'}|$. \ignore{We denote by
$\nit{MRes}(D,M)$ the set of minimally resolved instances of $D$ wrt
the set $M$ of MDs.}
\boxtheorem
\end{definition}
\begin{example} (Example \ref{ex:exp} continued) It holds $S_{D,D_1}= \{$ $(t_2,B), (t_4,B)\}$; and
$S_{D,D_2}$ $=$ $\{(t_2,B), (t_3,B), (t_4,B)\}$. Furthermore, $|S_{D,D_1}| < |S_{D,D_2}|$. \boxtheorem
\end{example}
The MRIs are the intended clean instances obtained after the application of a set
of MDs to an initial instance $D$. There is always an MRI for an instance $D$ wrt $M$ \cite{jaf11}.
The {\em clean} or {\em resolved}
answers to a query are  {\em certain} for the class of MRIs for $D$ wrt $M$. They are the intrinsically clean
answers to the query.

\begin{definition} \cite{jaf11}
Let $\mathcal{Q}(\bar{x})$ be a query expressed in the first-order
language $L(\mathcal{S})$ associated to schema $\mathcal{S}$ of an instance $D$.
 A tuple of constants $\bar{a}$ from $U$ is a {\em resolved answer} to $\mathcal{Q}(\bar{x})$ wrt the set
$M$ of MDs, denoted
$D \models_{M} \mathcal{Q}[\bar{a}]$, iff  $D' \models \mathcal{Q}[\bar{a}]$, for every MRI $D'$ of $D$ wrt $M$.
\ignore{If $\bar{x}$ is empty, i.e. $\mc{Q}$ is a boolean query, {\it true} is the resolved answer if $D' \models \mathcal{Q}$, for every MRI $D'$; otherwise, the resolved answer is {\it false}.}
We denote with $\nit{Res\!An}(D,\mathcal{Q},M)$ the set of resolved answers to $\mathcal{Q}$ from $D$ wrt $M$.
\boxtheorem
\end{definition}

\section{On the Complexity of RAP}\label{subsec:com}

Notice that the number of MRIs can be exponential in the size of the instance, as the next example shows.

\begin{example} (example \ref{ex:exp} continued) The example can be generalized with the following instance:
\begin{center}
\begin{tabular}{c|c|c|}\hline
$R(D^n)$&$A$ & $B$  \\ \hline
$t_1$&$a_1$ & $c_1$  \\
$t_2$&$a_1$ & $c_2$  \\
$\cdots$ & $\cdots$ & $\cdots$\\
$t_{2n-1}$&$a_n$ & $c_{2n-1}$  \\
$t_{2n}$&$a_n$ & $c_{2n}$ \\ \cline{2-3}
\end{tabular}
\end{center}
This instance
with $2n$ tuples has $2^n$ MRIs.
\boxtheorem
\end{example}
Checking the possibly exponentially many MRIs for an instance to obtain resolved answers is inefficient. We need more efficient algorithms. However, this aspiration will be limited by the intrinsic complexity
of the problem. In this work we investigate the complexity of
computing resolved answers to queries. We concentrate on the
{\em resolved answer  problem} (RAP), about deciding if a tuple is a resolved answer.

\begin{definition} \label{def:rap}
 For a query $\mathcal{Q}(\bar{x}) \in L(\mathcal{S})$, and  $M$, the {\em resolved answer problem} is deciding membership of the set:

\vspace{1mm}
\hspace*{1cm}$\nit{RA}_{\mathcal{Q},M} :=
\{(D,\bar{a})~|~\bar{a}\in \nit{Res\!An}(D,\mathcal{Q},M)\}$. \boxtheorem
\end{definition}
\ignore{Notice that when $\mathcal{Q}$ is
boolean, this decision problem becomes $\nit{RA}_{\mathcal{Q},M} =
\{D~|$ $\nit{true} \in \nit{Res\!An}(D,\mathcal{Q},M)\}$.}
A different decision problem, closely
related to RAP, was shown to be intractable when there is more
than one MD \cite{jaf11}. This is because new similarities
can arise between values as a result of a particular choice of update
values (rather than because the values were identified as duplicates
and merged). Such similarities are called {\em accidental similarities} \cite{jaf11}. As we will see, this dependence of updates on the choice of
update values for previous updates may  make RAP
intractable.

\begin{example} (Example \ref{ex:exp} continued)
For instance $D_2$,  a similarity for attribute  $B$ is ``accidentally" created for tuples $t_2, t_3$. \boxtheorem
\end{example}
Since duplicate resolution involves modifying
individual values,
an important problem is to decide which
of these values are the same in all MRIs. It is obviously related to the RAP problem, and sheds light on its
complexity. More precisely,
for a fixed predicate $R$, and $A$ an attribute of $R$ in position $i$, we consider the unary query \ $\mc{Q}^{R.A}(x_i)\!:$
\begin{equation} \label{eq:ra}
 \exists x_1 \cdots x_{i-1}x_{i+1}\cdots x_n R(x_1, \ldots, x_{i-1},x_i,x_{i+1},\ldots, x_n),
\end{equation}
i.e. the projection of $R$ on $A$; and a special case  of RAP:
\begin{equation}
\nit{RA}^{R.A}_M = \{(D,a) ~|~ a \in \nit{Res\!An}(D,\mathcal{Q}^{R.A}(x_i),M)\}.
\end{equation}
Intractability of simple {\em single-projected atomic queries} like (\ref{eq:ra}), i.e. of
$\nit{RA}^{R.A}_M$, restricts the general efficient applicability of duplicate
resolution. On the other hand, we will show (cf. Sections \ref{sec:Tr}, \ref{subsec:NI2}) that, for important classes of conjunctive
queries and for sets of MDs such that $\nit{RA}^{R.A}_M$ can be efficiently solved for all $R$ and
$A$, the resolved answers to queries in the class can be efficiently computed. For this reason, we concentrate
on the following classification
of MDs.

\begin{definition}\label{def:hardmd}
A set $M$ of MDs is
{\em hard} if, for some predicate $R$
and some attribute $A$ of $R$, $\nit{RA}^{R.A}_M$ is $\nit{NP}$-hard (in data).
\ $M$ is {\em easy} if, for each
$R$ and $A$, $\nit{RA}^{R.A}_M$ can be solved in polynomial time.\footnote{The problem used here to define hard/easy  is
slightly different from, and more appropriate than, the one used in \cite{jaf11}. Here hardness refers to {\em Turing reductions.}\ignore{That  was about
identifying the set of all resolved answers,
rather than that of determining whether or not an individual
answer to the query is a resolved answer. This definition
provides a closer correspondence with RAP
than the definition in \cite{jaf11}.}}
\boxtheorem
\end{definition}
In the next subsections, we develop syntactic criteria on MDs for easiness/hardness (cf. Theorems \ref{thm:combine},
\ref{thm:conv}, and Definition \ref{def:HSC}).
 Some of these complexity results will be
generalized in Section \ref{sec:Tr} to larger classes
of conjunctive queries.

\subsection{Acyclic MDs and a dichotomy result}

Non-interacting (NI) sets of MDs (cf. Section \ref{sec:pre}) are easy, due to the simple form of the MRIs, each of which can
be obtained with a single update. So, sets of duplicate values
can be identified simply by comparing pairs of tuples in the given instance,
to see if they satisfy the similarity relations. The minimality
condition implies that each such set of duplicate values must be updated to (one of)
the most frequently occurring value(s) among them.
 The simplest non-trivial case is a {\em linear pair} of two MDs.

\begin{definition}\label{def:lin}
A {\em linear pair} $M$ of MDs is such that \linebreak $\nit{MDG}(M)$
consists of the vertices $m_1$ and $m_2$ with
an edge from $m_1$ to $m_2$. The
linear pair is denoted by $(m_1,m_2)$.
\boxtheorem
\end{definition}
The case of linear pairs is non-trivial in the sense that
it can be hard (cf. Theorem \ref{thm:conv}).
In this section, we show that tractability for
linear pairs occurs when the form of the
MDs is such that it prevents accidental similarities generated
in one update from affecting subsequent updates (cf. Theorem \ref{thm:combine}).
Deciding whether or not a linear pair has this form is straightforward.
Although all results of this section are
stated for MDs involving two distinct predicates, they can
easily be extended to the case of single relation.\footnote{This is done
by treating the relation as two different relations with
identical tuples and attributes. For example, the condition
$S[A]\approx S[B]$ is interpreted as $S_L[A_L]\approx S_R[B_R]$.
All complexity results go through with minor modifications.}

\begin{example}\label{ex:ahardset}
Consider the following linear pair $(m_1,m_2)$ of MDs and instance: \vspace{-2mm}
\bea
m_1:~R[A] = S[E]\ra R[B]\doteq S[F],\nn\\
m_2:~R[B] = S[F]\ra R[C]\doteq S[G].\nn
\eea
\begin{center}
\begin{tabular}{l|c|c|c|} \hline
$R$&$A$ & $B$ & $C$  \\ \hline
$t_1$&$a$ & $c$ & $g$  \\
$t_3$&$b$ & $c$ & $e$  \\ \cline{2-4}
\end{tabular}~~~
\begin{tabular}{l|c|c|c|} \hline
$S$&$E$ & $F$ & $G$  \\ \hline
$t_2$&$a$ & $d$ & $h$  \\
$t_4$&$b$ & $f$ & $k$ \\ \cline{2-4}
\end{tabular}
\end{center}
Different instances can be produced with a single update,
depending on the choice of common value.
Two of those instances are: \vspace{-1mm}
\begin{center}\begin{tabular}{l|c|c|c|}\hline
$R'$ &$A$ & $B$ & $C$  \\ \hline
$t_1$&$a$ & $d$ & $g$  \\
$t_3$&$b$ & $c$ & $e$  \\ \cline{2-4}
\end{tabular}~~~
\begin{tabular}{l|c|c|c|}\hline
$S'$ &$E$ & $F$ & $G$  \\ \hline
$t_2$&$a$ & $d$ & $h$  \\
$t_4$&$b$ & $c$ & $k$ \\ \cline{2-4}
\end{tabular}

\vspace{2mm}
\begin{tabular}{l|c|c|c|}\hline
$R''$ &$A$ & $B$ & $C$  \\ \hline
$t_1$&$a$ & $c$ & $g$  \\
$t_3$&$b$ & $c$ & $e$  \\ \cline{2-4}
\end{tabular}~~~
\begin{tabular}{l|c|c|c|}\hline
$S''$ &$E$ & $F$ & $G$  \\ \hline
$t_2$&$a$ & $c$ & $h$  \\
$t_4$&$b$ & $c$ & $k$ \\ \cline{2-4}
\end{tabular}
\end{center}

\noindent These two updates lead to
different sets of tuples with
duplicate values for the $R[C]$ and $S[G]$ attributes
to be matched, $\{t_1,t_2\}$ and $\{t_3,t_4\}$
in the case of $R'$, and $\{t_1,t_2,t_3,t_4\}$
in the case of $R''$. In general, the
effect of the choice of update values for the
$R[B]$ and $S[F]$ attributes on subsequent updates for the $R[C]$
and $S[G]$ attributes leads to intractability.
Actually,  this linear pair will turn out to be hard (cf. below).

However, an easy set of MDs can be obtained by introducing
the similarity condition of $m_1$ into $m_2$: \vspace{-2mm}
\bea
&&m_1:~R[A] = S[E]\ra R[B]\doteq S[F],\nn\\
&&m_2':~R[A] = S[E]\w R[B] = S[F]\ra R[C]\doteq S[G].\nn
\eea
The accidental similarity between, for example, $t_2[F]$ in $S''$
and $t_3[B]$ in $R''$ cannot affect the update on the $R[C]$ and $S[G]$ attribute
values of these tuples, because the $S[E]$ attribute value of $t_2$ and
the $R[A]$ attribute value of $t_3$ are dissimilar.
In effect, the conjunct $R[A] = S[E]$
``filters out" the accidental similarities generated by application of
$m_1$, preventing them from affecting the update on the $R[C]$ and $S[G]$
attribute values.
\boxtheorem
\end{example}
In general, any linear pair $(m_1,m_2)$
for which the similarity condition of $m_1$ is included in
that of $m_2$ is easy \cite{jaf11}.
Although linear pairs $(m_1,m_2)$ are, in general, hard,
the previous example shows that they can be easy if all attributes
in $\nit{LHS}(m_1)$ also occur in $\nit{LHS}(m_2)$.
We now generalize this result showing that, when
all similarity operators are {\em transitive},
a linear pair can
be easy iff a subset of
the attributes of $\nit{LHS}(m_1)$ are in $\nit{LHS}(m_2)$.

Transitivity is not necessarily assumed for a similarity relation. In consequence, it deserves a discussion.
Transitivity in this case requires
that two dissimilar values cannot be similar to the same value.
This imposes  a restriction on accidental similarities, as the next  example shows, extending the set of tractable cases.

\begin{example}\label{ex:accidental}
Consider the pair $M$, and instance $D$, only part of which
is shown below. The {\em only} similarities are: $e\approx a$ and $e\approx i$.
So, $\approx$ is non-transitive.
\bea
&&m_1:~R[A]\approx S[E]\w R[B]\approx S[F]\ra R[C]\doteq S[G]\nn\\
&&m_2:~R[A]\approx S[G] \w R[C]\approx S[G] \w R[C]\approx R[E]\ra \nn\\
&& \hspace*{5.7cm} R[H]\doteq S[I]\nn
\eea
\begin{center}
\begin{tabular}{c|c|c|c|}\hline
$R(D)$&$A$ & $B$ & $C$ \\ \hline
$t_1$&$a$ & $b$ & $e$\\
$t_3$&$a$ & $c$ & $e$\\
$t_5$&$i$ & $j$ & $e$\\
$t_7$&$i$ & $k$ & $e$\\ \cline{2-4}
\end{tabular}
~~~~~~~~~~~~\begin{tabular}{c|c|c|c|}\hline
$S(D)$&$E$ & $F$ & $G$  \\ \hline
$t_2$&$a$ & $b$ & $a$  \\
$t_4$&$a$ & $c$ & $a$ \\
$t_6$&$i$ & $j$ & $i$ \\
$t_8$&$i$ & $k$ & $i$ \\ \cline{2-4}
\end{tabular}
\end{center}
 The  first MD requires an update of each pair in the set
$\{(t_l[C],t_{l+1}[G])~|~1\leq l\leq 7,~l\hbox{ odd}\}$
to a common value. If $e$ is chosen
as this value for all pairs, then all pairs of tuples,
one from $R$ and one from $S$, would satisfy the
similarity condition of $m_2$, causing the values
of $t[H]$ to be updated to a common value for all tuples
in $R$. However, if in the initial update $a$ is chosen as the update value
for $(t_1[C],t_2[G])$ and $(t_3[C],t_4[G])$, and $i$ is chosen as the
update value for $(t_5[C],t_6[G])$ and $(t_7[C],t_8[G])$,
then the value of $\{t_1[H],t_3[H]\}$ and that of
$\{t_5[H],t_7[H]\}$ will be updated independently of
each other. If $\approx$ were transitive, this would always
be the case, leaving fewer possibilities for updates.
\boxtheorem
\end{example}
Most similarity relations used in ER
are not transitive \cite{elmargamid}. While
this restricts the applicability of the tractability results
presented in this subsection, they could still be applied in situations
where the non-transitive similarity relations satisfy transitivity to a good approximation,
for the
specific instance at hand.

Consider Example \ref{ex:accidental}, assuming string-valued attributes, and $\approx$
defined as the property of being within a certain
{\em edit distance}, which is not transitive. Accidental similarities, such as the one in Example \ref{ex:accidental},
may arise in general.
However, one could expect
the edit distance between duplicate values within the
$R[A]$ column to be very
small relative to that between non-duplicate values.
This would be the case if errors were
small within those columns. In such a case, the edit distance threshold could be
chosen so that the duplicate values
would be clustered into groups of mutually similar values,
with a large edit distance between any two values from
different groups.

In Example \ref{ex:accidental},
if $a$ and $i$ are dissimilar,
the pair of similarities $e\approx a$ and $e\approx i$
that led to the accidental similarities when $e$ was chosen
as the update value would be unlikely to occur.
Since such accidental similarities, which are precluded
when $\approx$ is transitive, are rare in this case,
they would affect only a few tuples in the instance.
In consequence, a good approximation to the resolved answers would
be obtained by applying a polynomial time algorithm that
returns the resolved answers under the assumption that $\approx$ is transitive.
In this paper we do not investigate this direction any further. The easiness
results (but not the hardness results) presented in this section require
the assumption of transitivity of all similarity operators. They do not
hold in general
for non-transitive similarity relations.

\begin{definition}
Let $m$ be an MD. The symmetric binary
relation $\nit{LRel}_m$ ($\nit{RRel}_m$) relates each pair of attributes
$A$ and $B$ such that a conjunct of the form $R[A]\approx S[B]$ ($R[A]\doteq S[B]$)
appears in $\nit{LHS}(m)$ ($\nit{RHS}(m)$). An {\em L-component}
({\em R-component}) of $m$ is an equivalence
class of the reflexive, transitive closure, $\nit{LRel}_m^{eq}$ ($\nit{RRel}_m^{eq}$), of
$\nit{LRel}_m$ ($\nit{RRel}_m$).
\boxtheorem
\end{definition}

\begin{lemma}\label{thm:filter}
A linear pair $(m_1,m_2)$ of MDs, with $\approx_1$
and $\approx_2$ transitive,  and $R$, $S$ distinct relations,\vspace{-2mm}
\bea
m_1\!:~R[\bar A]\approx_1 S[\bar B]\ra R[\bar C]\doteq S[\bar E]\nn\\
m_2\!:~R[\bar F]\approx_2 S[\bar G]\ra R[\bar H]\doteq S[\bar I]\nn
\eea
is easy
if the following holds: If an attribute of $R$ ($S$) in $\nit{RHS}(m_1)$
occurs in $\nit{LHS}(m_2)$, then for each L-component $L$ of $m_1$, there is an
attribute of $R$ ($S$) from $L$ that belongs to $\nit{LHS}(m_2)$.
\boxtheorem
\end{lemma}

\begin{example}\label{ex:easyfilter} Assuming that $\approx$ is transitive,
the following linear pair of MDs: \vspace{-2mm}
\bea
m_1\!:&&\hspace*{-4mm}R[A]\approx S[B]\w R[C]\approx S[B]\w R[E]\approx S[F]\ra\nn\\
&&\hspace*{5cm} R[G]\doteq S[H],\nn\\
m_2\!:&&\hspace*{-4mm}R[G]\approx S[H]\w R[A]\approx S[B]\w R[E]\approx S[F] \ra \nn\\
&& \hspace*{5cm}R[I]\doteq S[J]\nn
\eea
 is easy, because Lemma \ref{thm:filter} applies.  Here, the L-components of $m_1$ are
$\{R[A],$ $R[C],S[B]\}$ and
$\{R[E],S[F]\}$. Here, \linebreak $\nit{LHS}(m_2)$ includes both an attribute
of $R$ and an attribute of $S$ from each of these
L-components. \boxtheorem
\end{example}
Lemma \ref{thm:filter} generalizes the idea of Example \ref{ex:ahardset}, where
with $(m_1,m_2')$, accidental similarities
are ``filtered out" and cannot affect updates.
In some cases, a linear pair of MDs can be easy despite
the presence of accidental similarities which can affect
subsequent updates. This happens when an attribute must
take on a specific value in order to affect further updates.
Definitions \ref{def:equivset} and \ref{def:bound}
syntactically capture this intuition. $\nit{TC}(r)$ denotes the
transitive closure of a binary relation $r$.

\begin{definition}\label{def:equivset}
Let $(m_1,m_2)$ be a linear pair of MDs of the form \vspace{-5mm}
\bea
m_1\!:~R[\bar A]\approx_1 S[\bar C]\ra R[\bar E]\doteq S[\bar F]\nn\\
m_2\!:~R[\bar G]\approx_2 S[\bar H]\ra R[\bar I]\doteq S[\bar J]  \nn
\eea
(a) For predicate $R$, $B_R$ is a binary relation on
attributes of $R$: For attributes $R[A_1]$ and $R[A_2]$,
$B_R(R[A_1],R[A_2])$ holds iff $R[A_1]$ and $R[A_2]$ are in the same
R-component of $m_1$ or the same L-component of $m_2$.
Relation $B_S$ is defined analogously for predicate $S$.

\noindent (b) An {\em equivalent set} (ES) of attributes of $(m_1,m_2)$
is an equivalence class of $\nit{TC}(B_R)$
or of $\nit{TC}(B_S)$, with at least one attribute in the equivalence class
belonging to $\nit{LHS}(m_2)$.
\boxtheorem
\end{definition}
Notice that relations $B_R$ and $B_S$ are reflexive and symmetric binary relations on attributes in
$\nit{RHS}(m_1)\cup \nit{LHS}(m_2)$.

\begin{example}\label{ex:equivset}
Consider the following linear pair of MDs on relations
$R[A,C,E,G,H]$ and $S[B,D,F,I]$: \vspace{-2mm}
\bea
R[A]\approx S[B]\ra R[C]\doteq S[D]\w R[E]\doteq S[D]\nn\\
R[E]\approx S[F]\w R[G]\approx S[F]\ra R[H]\doteq S[I]\nn
\eea
The attributes of $R$ satisfy the relations $B_R(R[C],R[E])$
(due to $R[C]\doteq S[D]$ and
$R[E]\doteq S[D]$) and $B_R(R[E],R[G])$
(due to $R[E]\approx S[F]$ and $R[G]\approx S[F]$).
Relation $B_S$ is empty, since there is only
one attribute of $S$ in each of $\nit{RHS}(m_1)$ and $\nit{LHS}(m_2)$.
There is one non-singleton ES, $\{R[C],R[E],R[G]\}$, and
also the singleton ES $\{S[F]\}$.
\boxtheorem
\end{example}
An ES is a natural unit that  groups together the attributes of a linear pair with
transitive similarities, because of the close association between the update
values for them. For a linear pair
as in Definition \ref{def:equivset}, the set of values which a tuple $t$ in relation $R$
takes on the attributes within an R-component of $m_1$ must be modified to the same value if
any of the values is modifiable. Also, by transitivity, the attributes of
$t$ in $\nit{RHS}(m_2)$ are not modifiable by $m_2$ unless the values taken by $t$ on
the attributes in an L-component of $m_2$ are similar (cf. Example \ref{ex:natural} below).
Therefore, when considering updates that affect the values of
attributes in $\nit{RHS}(m_2)$, the values for a given tuple of attributes
within an ES of attributes can be assumed to be similar.

\begin{example}\label{ex:natural}
(example \ref{ex:accidental} continued)
We illustrate the association between
values of attributes in an ES, and also
 how the presence of an ES of a certain
form can simplify updates.

With the given instance and set $M$ of MDs, we now
assume that $\approx$ is transitive. $M$ has
the ES $\{R[A],$ $R[C]\}$. For any tuple $t$
of $R$, the value of $t[A]$ must be similar to that of $t[C]$ in order for
there to be a tuple $t'$ in $S$ such that $t$ and $t'$ satisfy the similarity
condition of $m_2$. This is because they must
both be similar to the value of $t'[G]$, and then must be similar
to each other by transitivity. If there is no such tuple $t'$,
then by Definition \ref{def:mod}, $t[H]$ is not modifiable, and
by Definition \ref{def:new}, the value of $t[H]$ does not change.

$M$  does
not satisfy the condition of Lemma \ref{thm:filter}. Here, unlike those for which Lemma \ref{thm:filter} holds,
the application of the MDs
can result in accidental similarities between pairs of modifiable values
in $R$ that can affect further updates. This is because only
$R[A]$, not both $R[A]$ and $R[B]$, is in $\nit{LHS}(m_2)$ (cf.  Lemma \ref{thm:filter}).
For example, when $m_1$ is applied to
the instance, if both the pair $t_1[C]$ and $t_2[G]$, and the pair $t_3[C]$
and $t_4[G]$ are updated to $a$, there will be an accidental
similarity between $t_1[C]$ and $t_3[C]$, forcing to update $t_1[H]$
and $t_3[H]$ to a common value.

Despite these accidental similarities, updates are made simpler by the
fact that the ES contains $R[A]$,
an attribute in $\nit{LHS}(m_1)$. All sets of tuples
in $R$ whose values for $R[C]$ are matched must
have the same value for $R[A]$. After these values
are merged, regardless of the
common value chosen, either
all tuples in the set will have their $R[H]$ values
changed, or none of them will change. This would not
be true in general if there were no attribute of $\nit{LHS}(m_1)$
in the ES. In that case, there could be
many possible outcomes depending on the value chosen for
a set of duplicate values of $R[C]$.
\boxtheorem
\end{example}
Example \ref{ex:natural} shows how, for a linear pair $(m_1,m_2)$,
the presence of an attribute of $\nit{LHS}(m_1)$ in an ES can simplify updates. This motivates the next definition.

\begin{definition}\label{def:bound}
Let $(m_1,m_2)$ be a linear pair of MDs on relations
$R$ and $S$. An ES $E$ of $(m_1,m_2)$ is {\em bound} if
$E\cap  \nit{LHS}(m_1)$ is non-empty.
\boxtheorem
\end{definition}
\begin{example}\label{ex:bound}
Consider the following linear pair of MDs defined on  $R[A,C,F,H,I,M]$ and $S[B,D,E,G,N]$:
\bea
&& R[A]\approx S[B] \ra R[C]\doteq S[D]\w\nn\\
&& \hspace*{1cm}R[C]\doteq S[E]\w R[F]\doteq S[G]\w R[H]\doteq S[G],\nn\\
&& R[F]\approx S[E]\w R[I]\approx
S[E]\w R[A]\approx S[E]\w\nn\\
&& \hspace*{2.9cm}R[F]\approx S[B]\ra R[M]\doteq S[N].\nn
\eea
The ES $\{S[D],S[E],S[B]\}$ is bound,
because it contains $S[B]$. The ES
$\{R[A],R[F],R[I],R[H]\}$ is bound, because it contains
$R[A]$.
\boxtheorem
\end{example}

\begin{lemma}\label{thm:bind}
A linear pair $(m_1,m_2)$ of MDs as in
Lemma \ref{thm:filter} is easy if all ESs are bound.
\boxtheorem
\end{lemma}

\begin{example}\label{ex:easybind} (examples \ref{ex:accidental} and \ref{ex:natural} continued)
If $\approx$ is transitive, it follows from Lemma \ref{thm:bind} that $M$  in Example \ref{ex:accidental} is easy. As we verified in Example \ref{ex:natural},  $M$
 does not satisfy the conditions of Lemma \ref{thm:filter}. \boxtheorem
\end{example}
$M$
of Example \ref{ex:accidental} does not satisfy the conditions of Lemma \ref{thm:filter}, but satisfies
those of Lemma \ref{thm:bind}.
On the other hand, $M$ of Example \ref{ex:easyfilter} satisfies the conditions of
Lemma \ref{thm:filter}, but not those of
Lemma \ref{thm:bind}.  However, $M$ of Example \ref{ex:bound}
satisfies both. This shows that the two easiness conditions
are independent, but not mutually exclusive. Actually,
Lemmas \ref{thm:filter} and \ref{thm:bind} combined give us the
following result, which subsumes each of them.

\begin{theorem}\label{thm:combine}
Let $(m_1,m_2)$ be a linear pair  as in
Lemma \ref{thm:filter}. For predicate $R$, let $E_R$ be
the class of ESs of $(m_1,m_2)$ that
are equivalence classes of  $\nit{TC}(B_R)$. $E_S$ is
defined similarly using $B_S$.\footnote{Thus, elements of $E_R$
are ESs in the sense of Definition \ref{def:equivset}(b), but for
$\nit{TC}(B_R)$ as opposed to $\nit{TC}(B_R) \cup \nit{TC}(B_S)$.} \ $(m_1,m_2)$ is easy
if both of the following hold:\\
(a) At least one of the following is true: (i) there are no
attributes of $R$ in $\nit{RHS}(m_1)\cap\nit{LHS(m_2)}$; (ii) all ESs in $E_R$ are bound; or (iii)
for each L-component $L$ of $m_1$, there is an attribute
of $R$  in $L \cap \nit{LHS}(m_2)$.\\
(b) At least one of the following is true: (i) there are no
attributes of $S$ in $\nit{RHS}(m_1)\cap\nit{LHS(m_2)}$; (ii) all ESs in $E_S$ are bound; or (iii)
for each L-component $L$ of $m_1$, there is an attribute
of $S$ in $L \cap \nit{LHS}(m_2)$.
\boxtheorem
\end{theorem}
In the rest of this section, we will obtain a partial converse of
Theorem \ref{thm:combine}. For this purpose,
 we make the assumption that, for each similarity relation,
there is an infinite set of mutually dissimilar
elements. Strictly speaking, the results below require only that
the set of mutually dissimilar elements be at least as large
as any instance under consideration. This is assumed in our
next hardness result for certain linear pairs. We expect this
assumption to be satisfied by many similarity measures used
in practice, such as the edit distance and related
similarities based on string comparison.

The proof is by polynomial reduction from a decision problem
that we call {\em Cover Set} (CS) that is related to the well-known {\em minimum set-cover} (MSC).
Given $\mc{I} = \langle \mc{U}, \mc{C}, S\rangle$, with
$\mc{U}$ is a set, $\mc{C}$ a collection
of subsets of $\mc{U}$ whose union is $\mc{U}$, and $S \in \mc{C}$, the problem
is deciding whether or not there is a minimum (cardinality) set cover $\mc{S}'$ for $\langle \mc{U},\mc{S}\rangle$ with
$S \in \mc{S}'$.
This problem is $\nit{NP}$-complete.\footnote{Cf. Lemma \ref{lem:minset}
in the appendix.}
The reduction constructs a finite database instance $D$, where every pair of values in it that are different
are also dissimilar. However, a value may appear more than once. Certain values in $D$ are associated with elements
of $\mc{U}$ or $\mc{C}$. This reduction is indifferent to whether
or not the similarity relations are transitive, since distinct values
in the instance are dissimilar, and equal values are similar by equality
subsumption.

\begin{theorem}\label{thm:conv} Assume  each similarity relation has an infinite
set of mutually dissimilar elements. Let $(m_1,$ $m_2)$ be a
linear pair of MDs with
$\nit{RHS}(m_1) \cap \nit{RHS}(m_2) = \ems$.
If $(m_1,m_2)$ does not satisfy the condition of Theorem \ref{thm:combine},
then it is hard.\footnote{The assumption
$\nit{RHS}(m_1)\cap \nit{RHS}(m_2) = \ems$ is used to ensure that
a resolved instance is always obtained after a fixed number of updates
(actually two), making it easier to restrict the form MRIs can take. This is used
in the hardness proofs.}
\boxtheorem
\end{theorem}

\begin{example}\label{ex:hardcases}
We can apply Theorem \ref{thm:conv} to identify hard sets of MDs. (Assuming
for each  similarity relation involved an infinite set of mutually dissimilar elements.)

The set of MDs in Example \ref{ex:ahardset} is hard, because condition
(a) of Theorem \ref{thm:combine} does not hold, because all of the following hold: (i) there is an attribute,
$R[B]$ of $R$, in $\nit{RHS}(m_1)\cap\nit{LHS(m_2)}$; (ii) the ES $\{R[B]\}$ is not bound; and (iii) there is no attribute of $R$ in the
L-component $\{R[A],$ $S[E]\}$ that belongs to $\nit{LHS}(m_2)$.

The
set of MDs in Example \ref{ex:accidental} is hard, because condition (b)
of Theorem \ref{thm:combine} does not hold, because all of the following hold: (i) there is an attribute,
$S[E]$ of $S$, in $\nit{RHS}(m_1)\cap\nit{LHS(m_2)}$; (ii) the ES $\{S[E]\}$ is not bound; and (iii) there is no attribute of $S$ in the
L-component $\{R[A],$ $S[C]\}$ that belongs to $\nit{LHS}(m_2)$.

The set
of MDs in Example \ref{ex:equivset} is hard, because condition (a) of
Theorem \ref{thm:combine} does not hold, because all of the following hold: (i) there are attributes
of $R$ in $\nit{RHS}(m_1)\cap\nit{LHS(m_2)}$; (ii) the ES
$\{R[C], R[E], R[G]\}$ is not bound; and (iii) there is no attribute
of $R$ in the L-component $\{R[A],$ $S[B]\}$ that belongs to $\nit{LHS}(m_2)$.
\boxtheorem
\end{example}
Theorem \ref{thm:conv} does not require the transitivity of the similarity relations, which is
needed for tractability. Theorems \ref{thm:combine} and \ref{thm:conv}
imply the following {\em dichotomy} result. It tells us that for a syntactic class of linear pairs, each
of its elements is easy or hard. That is, there is nothing ``in between", which is not necessarily true
in general. Actually, if $P \neq \nit{NP}$, there are decision problems in \nit{NP} between $P$ and $\nit{NP}$-complete \cite{ladner}.

\begin{theorem}\label{thm:dich} Assume each similarity relation is transitive and has  an infinite
set of mutually dissimilar elements. Let $(m_1,m_2)$ be a linear pair
of MDs with $\nit{RHS}(m_1) \cap \nit{RHS}(m_2)$ $= \ems$. Then,
$(m_1,m_2)$ is either easy or hard.
\boxtheorem
\end{theorem}
Theorem \ref{thm:dich} divides the class of linear pairs satisfying
certain conditions into an easy class, and a hard one. Deciding the membership
of either of them requires a simple syntactic checking procedure.
The dichotomy result shows that
very simple pairs of MDs, even ones such as $m_1$ and $m_2$ in Example \ref{ex:ahardset},
with equality as similarity, are hard.

Given the high computational complexity
of RAP for sets of two MDs, an important question is whether or not
larger sets of interacting MDs can be easy. We provide a positive answer
to this question in the next subsection. In the rest of the paper, we do not
assume transitivity of similarity relations.

\subsection{Cyclic sets of MDs}\label{subsec:cycle}

We described above how acyclic sets
of MDs can be easy if the possible effects of accidental similarities
are restricted. Here, we present
a different class of easy sets of MDs for which
such effects are not
restricted. Actually, we establish the somewhat
surprising result that certain cyclic sets of MDs
are easy. In this section we do not make the assumption that
each MD involves different predicates.

\begin{definition}\label{def:scycle}
A set $M$ of MDs is {\em simple-cycle} (SC) if its MD graph $\mdg$ is
(just) a cycle, and: \
(a) in all MDs in $M$ and in all their corresponding pairs,
the two attributes (and predicates) are the same; and \
(b) in all MDs $m \in M$, at most one attribute
in $\nit{LHS}(m)$ is changeable. \boxtheorem
\end{definition}

\ignore{\comlb{When you say above "is a cycle", you mean just a single cycle and nothing more?}\\
\comj{yes}
}
\begin{example}\label{ex:simple}
For schema  $R[A,C,F,G]$, consider the following set $M$ of MDs: \vspace{-2mm}
\bea
&&m_1\!: \ R[A]\approx R[A]\ra R[C,F,G]\doteq R[C,F,G],\nn\\
&&m_2\!: \ R[C]\approx R[C] \ra R[A,F,G]\doteq R[A,F,G].\nn
\eea
$\mdg$ is a cycle, because the attributes in $\nit{RHS}(m_2)$ appear in
$\nit{LHS}(m_1)$, and vice-versa. Furthermore, $M$ is SC, because
each of $\nit{LHS}(m_1)$ and $\nit{LHS}(m_2)$ are singletons.
\boxtheorem
\end{example}
For SC sets of MDs,
it is easy
to characterize the form taken by an MRI.

\begin{example}\label{ex:cycle}
Consider the  instance $D$ and a SC set of MDs, where the only similarities are: \
$a_i \approx a_j, \
b_i \approx b_j, \ d_i \approx d_j, \ e_i \approx e_j$, with $i,j \in
\{1,2\}$.
\vspace{-2mm}
\begin{multicols}{2}
\begin{tabular}{c|c|c|} \hline
$R(D)$&$A$ & $B$   \\ \hline
1&$a_1$ & $d_1$   \\
2&$a_2$ & $e_2$   \\
3&$b_1$ & $e_1$   \\
4&$b_2$ & $d_2$  \\ \cline{2-3}
\end{tabular}

\hspace*{-1.5cm}$m_1\!:~R[A]\approx R[A]\ra R[B]\doteq R[B]$,

\hspace*{-1.5cm}$m_2\!:~R[B]\approx R[B]\ra R[A]\doteq R[A]$.

\vspace{1mm}
\hspace*{-1cm} \noindent If the MDs are applied twice, \linebreak
\hspace*{-1cm}successively, starting from $D$,  a \linebreak
\hspace*{-1cm}possible result is:
\end{multicols}

\vspace{-2mm}
\begin{center}
\begin{tabular}{c|c|c|} \hline
$R(D)$&$A$ & $B$     \\ \hline
1&$a_1$ & $d_1$   \\
2&$a_2$ & $e_2$    \\
3&$b_1$ & $e_1$    \\
4&$b_2$ & $d_2$   \\ \cline{2-3}
\end{tabular}
~~~$\ra$~~~
\begin{tabular}{c|c|c|} \hline
$R(D_1)$&$A$ & $B$  \\ \hline
1&$b_2$ & $d_1$ \\
2&$a_2$ & $d_1$ \\
3&$a_2$ & $e_1$   \\
4&$b_2$ & $e_1$  \\ \cline{2-3}
\end{tabular}
\end{center}

\begin{center}
~~~~~~~~~~~~~~~~~~~~~~~~~~~~$\ra$~~~
\begin{tabular}{c|c|c|} \hline
$R(D_2)$&$A$ & $B$  \\ \hline
1&$a_2$ & $e_1$ \\
2&$a_2$ & $d_1$ \\
3&$b_2$ & $d_1$  \\
4&$b_2$ & $e_1$  \\ \cline{2-3}
\end{tabular}
\end{center}
\noindent It should be clear that,
in any sequence of instances
$D_1,D_2,$ $\ldots$, obtained from $D$ by
applying the MDs, the updated
instances must have the following pairs
of values equal (shown through the tuple ids):

\vspace{-2mm}
{\small \begin{table}[h]
\begin{center}
\begin{tabular}{|c|c|c|}\hline
$D_i$ \ \  $i$ odd & $A$ & $B$    \\ \hline
tuple (id) pairs & $(1,4)$, $(2,3)$ & $(1,2)$, $(3,4)$ \\ \hline
\end{tabular}
~~~~~
\begin{tabular}{|c|c|c|}\hline
$D_i$ \ \ $i$ even & $A$ & $B$    \\ \hline
tuple (id) pairs & $(1,2)$, $(3,4)$ & $(1,4)$, $(2,3)$ \\ \hline
\end{tabular}
\caption{Table of matchings}\label{table:changes}
\end{center}
\end{table}
}

\vspace{-4mm}\noindent In any stable instance, the pairs of values in the
above tables must be equal. Given the alternating behavior, this can only be
the case if all values in  $A$ are equal, and similarly for $B$, which can be achieved with a single
update, choosing any value as the common value for each of $A$ and $B$. In particular, an
 MRI requires the common value for each attribute to be set to a most common
value in the original instance. For $D$ there are 16
MRIs.

Set $M$ is easy: For any given instance $D$,
a table like Table \ref{table:changes} can be constructed, and using it, the sets of
duplicate values (i.e. values that are different, but should be equal) in the $R[A]$ and $R[B]$ columns can be matched in quadratic time. Given those
sets of duplicate values, and without having to actually match them, the resolved answers to the (single-projected
atomic) queries $\ex y R(x,y)$ and
$\ex x R(x,y)$ can be obtained from those values that occur within a (possibly singleton) set of duplicates
more often than any other value. For instance $D$, these queries return
the empty set.
\boxtheorem
\end{example}

\begin{proposition}\label{prop:cycle}
 Simple-cycle sets of MDs are easy.
\boxtheorem
\end{proposition}
The proof of this proposition can be done directly using an argument such
as the one given for Example \ref{ex:cycle}. However, this result
will be subsumed by a similar one for a broader class
of MDs (cf. Definition \ref{def:HSC}). SC sets of MDs can be easily found in practical applications.

\begin{example}\label{ex:practical}
(example \ref{ex:simple} continued) The relation $R$ subject to the given $M$,
has two ``keys", $R[A]$ and $R[C]$. A
relation like this may appear in a database
about people: $R[A]$ could be used for the person's
name, $R[C]$ the address, and $R[F]$ and $R[G]$ for non-distinguishing
information, e.g. gender and age. Easiness of $M$
can be shown  as in Example \ref{ex:cycle}, and also follows from Proposition
\ref{prop:cycle}. \boxtheorem
\end{example}
We show easiness for an extension of the class of SC MDs.

\begin{figure}
\centering
\epsfig{file=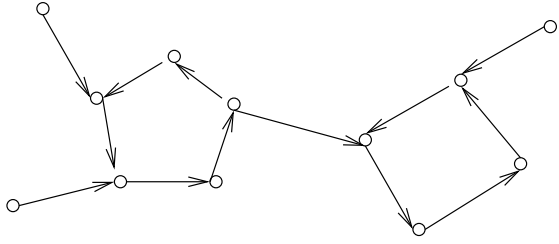}
\caption{The MD-graph of an HSC  set of MDs}\label{fig2}
\end{figure}

\begin{definition} \label{def:HSC} A  set
$M$ of MDs with MD-graph $\mdg$ is {\em hit-simple-cyclic} (HSC) iff:\\
(a) $M$ satisfies conditions (a) and (b) in
Definition \ref{def:scycle}; and\\
(b) each vertex $v_1$
in $\mdg$ is on at least one cycle or is connected
to a vertex $v_2$ on a cycle of non-zero length by an edge
directed toward $v_2$. \boxtheorem
\end{definition}
Notice that SC sets are also HSC sets.
An example of the MD graph of an HSC set of MDs is shown in Figure \ref{fig2}.

As the previous examples suggest, it is possible to provide
a full characterization of the MRIs for an instance subject
to an HSC set of MDs, which  we do next. It
will be used to prove that HSC sets of MDs are easy
(cf. Theorem \ref{prop:HSCeasy}). For this result, we
need a few definitions and notations.

For an SC set $M$ and  $m \in M$, if a pair of tuples satisfies
the similarity condition of {\em any} MD in $M$, then
the values of the attributes in $\nit{RHS}(m)$ must be merged for
these tuples.
Thus, in Example \ref{ex:cycle}, a pair of tuples satisfying {\em either}
$R[A]\approx R[A]$ {\em or} $R[B]\approx R[B]$ have {\em both}
their $R[A]$ and $R[B]$ attributes updated to the same value.
More generally, for an HSC set $M$ of MDs, and  $m \in M$, there is only a
{\em subset} of the MDs such that, if a pair of tuples satisfies
the similarity condition of an MD in the subset, then the
values of the attributes in $\nit{RHS}(m)$ must be merged for
the pair of tuples. We now formally define this subset.

\begin{definition}
Let $M$ be a set of MDs, and $m\in M$. The
{\em previous set} of $m$, denoted $PS(m)$, is the
set of all MDs $m'\in M$ with a path in
$\mdg$ from $m'$ to $m$.
\boxtheorem
\end{definition}
When applying a set of MDs to an instance, consistency
among updates made by different MDs must be enforced. This generally requires
computing a transitive closure relation that involves both
a pair of tuples and a pair of attributes.
For example, suppose $m_1$ has the conjunct $R[A]\doteq S[B]$ and $m_2$
has the conjunct $R[C]\doteq S[B]$. If $t_1$ and $t_2$ satisfy the condition
of $m_1$, and $t_2$ and $t_3$ satisfy the condition of $m_2$, then
 $t_1[A]$ and $t_3[C]$ must be updated to the same value, since updating
them to different values would require $t[B]$ to be updated to two different
values at once.
 We formally define this relation.\footnote{This  relation is actually more general than needed for HSC sets of MDs,
since each corresponding pair has the same attributes. However,
the more general case is needed when discussing
NI sets of MDs.}

\begin{definition}\label{def:taclosure} Consider an instance $D$, and $M = \{m_1,m_2,$ $\ldots, m_n\}$, with \vspace{-3mm}
\bea
m_i\!:~R[\bar A_i]\approx_i S[\bar B_i]\ra R[\bar C_i]\doteq S[\bar E_i].\nn
\eea
(a)  For $t_1, t_2 \in D$, $(t_1,C_i)\approx'(t_2,E_i)
\ :\Leftrightarrow$ \ $t_1[\bar{A}_j] \approx_j t_2[\bar{B}_j]$, where
$(C_i,E_i)$ is a corresponding pair of $(\bar C_i,\bar E_i)$
in $m_i$ and $m_j\in PS(m_i)$.
(b) The {\em tuple-attribute closure} of $M$ wrt $D$, denoted $\nit{TA}^{M,D}$, is
the reflexive, transitive closure of $\approx'$.
\boxtheorem
\end{definition}
Notice that $\approx'$ and $\nit{TA}^{M,D}$ are binary relations on
tuple-attribute pairs. To keep the notation simple, we will omit parentheses delimiting
tuple/attribute pairs in
elements of  $\nit{TA}^{M,D}$ (simply written as $\nit{TA}$). For example, for
tuples $t_1 = R(a,b,c)$ and $t_2 = S(d,e,f)$, with attributes $A,
C$ for $R, S$, resp., $\nit{TA}((t_1,A),(t_2,C))$ is simply written as $\nit{TA}(t_1,A,$ $t_2,C)$; and
similarly,  $\nit{TA}(((a,b,c),A),((d,e,f),C))$ as $\nit{TA}(a$ $,b,c,A,d,e,f,C)$.

In the case of NI and HSC sets of MDs, the MRIs for a given instance
can be characterized simply using the tuple/attribute closure.
This result is stated formally below.

\begin{proposition}\label{prop:HSC}
For $M$  NI or HSC, and $D$ an instance, each
MRI for $D$ wrt $M$ is obtained by setting, for each
equivalence class $E$ of $\nit{TA}^{M,D}$,
the value of all $t[A]$ for $(t,A)\in E$
to one of the most frequent values for $t[A]$ in $D$.
\boxtheorem
\end{proposition}

\begin{example}\label{ex:run} (Example \ref{ex:cycle} continued)
In this example, we represent tuples by their ids.
We have

\vspace{1mm}
$\nit{TA}^{M,D} = \{(i,A,j,A)~|~1\leq i,j\leq 4\}\ \cup$

\hspace*{4.3cm}$\{(i,B,j,B)~|~1\leq i,j\leq 4\}$,

\noindent whose equivalence classes are $\{(i,A)~|~1\leq i\leq 4\}$
and $\{(i,B)~|~1\leq i\leq 4\}$. From Proposition \ref{prop:HSC}
and the requirement of minimal change,
the 16 MRIs are obtained by setting all $R[A]$ and $R[B]$ attribute
values to one of the four existing  (and, actually, equally frequent) values for them.
\boxtheorem
\end{example}
 Proposition \ref{prop:HSC} implies that
for NI and HSC sets of MDs,
the set $E$ of sets of positions in an instance whose values are merged
to produce an MRI is the same for all MRIs (but the
common values chosen for them may differ, of
course). This
does not hold in general for arbitrary sets of MDs. Moreover, $E$
can be computed by taking the transitive closure of a binary relation
on values in
the instance, an $O(n^2)$ operation where $n$ is the size of the
instance. Given $E$, the resolved answers to the query $\mathcal{Q}^{R.A}$
are obtained as follows. For a tuple $t$ and attribute $A$,
the value $v$, with $t[A] = v$, is a resolved answer iff
for the equivalence class $S$ of $\nit{TA}$ to which $(t,A)$ belongs, for any
$v'\neq v$,
$|\{(t',B)\in S~|~t'[B] = v\}| > |\{(t',B)\in S~|~t'[B] = v'\}|$.
These observations lead to the following result.

\begin{theorem}\label{prop:HSCeasy}
HSC and NI sets of MDs are easy.
\boxtheorem
\end{theorem}
Theorem \ref{prop:HSCeasy}, does not imply that the set of all MRIs
can be efficiently computed. Because there can be $O(n)$ choices of
update value for each equivalence class of tuple/attribute closure, and
$O(n)$ such equivalence classes, there can be exponentially many
MRIs.

It may seem counterintuitive that HSC sets are
easy in light of the
fact that analogous non-cyclic cases such as the linear
pair $(m_1,m_2)$ of Example \ref{ex:ahardset} are hard. Indeed, while
tractability occurs in non-cyclic cases when accidental
similarities are ``filtered out" and cannot affect
the duplicate resolution process, cyclic cases are
easy for the opposite reason: all possible accidental
similarities are imposed on the values as these
similarities are propagated to all attributes in
the MDs on the cycle. Thus, the intractability
arising from having to choose common values so
as to avoid certain accidental
similarities is removed.

The tuple/attribute closure of Definition \ref{def:taclosure} can be defined using
a Datalog program, which we can use for query rewriting (cf. Section \ref{sec:Tr}).
Let $M$ be as in  Definition \ref{def:taclosure}.
Without losing generality and to simplify the presentation, we will assume in the rest of this section
 that predicates $R$ and $S$ are the same, so that we can keep them implicit.

The facts of the Datalog program, $\Pi^\nit{\!TA}_{\!D}$, are the ground atoms
$R(\bar a)$ in the original instance $D$, plus the facts of the form
$\bar c\approx_i \bar d$, that capture the similarity,
in the sense of $\approx_i$, of a pair of tuples $\bar c$ and $\bar d$
occurring in $D$.
Furthermore, $\Pi^\nit{\!TA}_{\!D}$ contains, for
each $m_i\in M$, for each corresponding pair
$R[A]\doteq R[B]$ in $m_i$, and for each $m_j \in \nit{PS}(m_i)$,
the rule \vspace{-2mm}
\bea
(\bar x,A)\approx' (\bar y,B)\lar R(\bar x), R(\bar y),
 \ \bar x\approx_j \bar y.\nn
\eea
The tuple/attribute closure $\nit{TA}^{M,(\cdot)}$ is given in Datalog as \vspace{-2mm}
\bea
&&\nit{TA}(\bar x,A,\bar y,B)\lar (\bar x,A)\approx' (\bar y,B).\nn\\
&&\nit{TA}(\bar x,A,\bar z,C)\lar \nit{TA}(\bar x,A,\bar y,B), \
(\bar y,B)\approx' (\bar z,C).\nn
\eea
Is it easy to verify that this program is finite and positive;  and that all its rules are {\em safe},
in the sense that all variables
appear in positive body atoms. The single minimal model of the program can be computed bottom-up,
as usual. This model captures the sets of value positions
to be merged which, as pointed out previously, are the
same for all MRIs of an instance to which a NI
or HSC set of MDs applies.

\begin{example}\label{ex:datalog}
(examples \ref{ex:cycle} and \ref{ex:run} continued)
For the MDs and instance of Example \ref{ex:cycle},
the facts of $\Pi^\nit{\!TA}_{\!D}$ are
$1\approx_1 2$, $3\approx_1 4$, $1\approx_2 4$, and
$2\approx_2 3$, where $\approx_i$ denotes the similarity
condition of $m_i$, in addition to the ground atoms in
$D$. Applying $\Pi^\nit{\!TA}_{\!D}$ gives
$(i,A)\approx'(i\mod 4 + 1,A)$ and
$(i,B)\approx'(i\mod 4 + 1,B)$, $1\leq i\leq 4$.
Applying the rule for $\nit{TA}$ we reobtain the classes  in Example \ref{ex:run}.
\boxtheorem
\end{example}
This
suggests a declarative specification of the resolved
answers: Given a conjunctive query, the query is rewritten by incorporating
the Datalog rules above. The combination retrieves the resolved
answers to the original query.  In the next section,
we will develop this approach for both NI
and HSC sets of MDs, to rewrite a query into one that retrieves
 the resolved answers to the original query. We will be able to provide
both a query rewriting methodology, and also an
extension of the
tractability results of this section (that refer to single-projected atomic queries) to a wider class of
conjunctive queries.

In this section we presented an algorithm that, taking
as input an instance $D$ and an HSC set
of MDs,  identifies the sets of duplicates
(i.e. sets of values that have to be matched) in time $O(n^2)$,
with $n = |D|$. This entails the easiness of
such sets of MDs (cf. Theorem \ref{prop:HSCeasy}).
We also introduced a Datalog program that can be used
to identify the duplicate sets, as an alternative to updating
the instance. The algorithm for duplicate set identification
can be easily extended into one that computes the
set of all MRIs for a given instance $D$. As expected,
the combination of the choices of common values may lead
to an exponential number of MRIs for $D$.

\section{Resolved Query Answering\ignore{:\\ Tractability and Rewriting}}\label{sec:Tr}

Here, we consider the two classes of easy sets
of MDs: NI and HSC sets of MDs. We will
 take advantage of the results of
 Section \ref{subsec:cycle}, to  efficiently retrieve the
resolved answers to queries in the \nit{UJCQ} class of conjunctive queries (cf. Definition
\ref{def:ujcq}). It extends the single-projected atomic queries (\ref{eq:ra}), which have a tractable RAP, by Theorem \ref{prop:HSCeasy}.

More precisely, we identify and discuss  tractable cases
of $\nit{RA}_{\mathcal{Q},M}$ for HSC and NI sets of MDs, and a certain class of conjunctive queries
 $\mc{Q}$. Actually, we present a
{\em query rewriting technique} for obtaining their resolved
answers. It works as follows. Given an instance $D$
and a query $\mathcal{Q}$, the
 MRIs for $D$ are not explicitly computed. Instead, $\mathcal{Q}$ is rewritten
into a new query $\mathcal{Q}'$, using both
$\mathcal{Q}$ and $M$. Query $\mc{Q}'$ is such that
when posed to $D$ (as usual), it returns the resolved answers to $\mathcal{Q}$ from $D$. \ $\mc{Q}'$ may not be a conjunctive query anymore. However,  if it can be efficiently evaluated against $D$,
the resolved answers can also be efficiently computed.\footnote{FO query rewriting  was applied in CQA, already in \cite{Arenas99} (cf. \cite{bertossi11} for a
survey)}. In our case, the rewritten queries will be (positive) Datalog queries
with aggregation (actually, $\nit{Count}$). They  can be
evaluated in polynomial time, making $RA_{\mathcal{Q},M}$ tractable.

The queries $\mc{Q}$ will be conjunctive, without built-in atoms, i.e. of
the form \
$\mathcal{Q}(\bar x)\!: \ \ex\bar u (R_1(\bar v_1) \wedge \cdots \wedge R_n(\bar v_n))$,
with $R_i \in \mc{R}$, and $\bar{x} = (\cup \bar{v_i}) \smallsetminus \bar{u}$.
Some additional restrictions on the joins we will be imposed below, to guarantee
the tractability of $RA_{\mathcal{Q},M}$.

\begin{definition} \label{def:ujcq}
Let $\mathcal{Q}$ be a conjunctive query,
and $M$ a set of MDs. Query $\mathcal{Q}$
is an {\em unchangeable join} conjunctive query
\ignore{($\nit{ucajCQ}$)} if there are no existentially quantified variables in a join
in $\mathcal{Q}$ in the position of a
changeable attribute. $\nit{UJCQ}$ denotes this class of queries. \boxtheorem
\end{definition}

\begin{example} For schema $\mc{S} = \{R[A,B]\}$, let $M$ consist of the single MD
$R[A]\approx R[A] \ra R[B]\doteq R[B]
$. Attribute $B$ is changeable, and $A$ is unchangeable. The query
$\mathcal{Q}_1(x,z)\!: \ \ex y (R(x,y)$ $\w$ $R(z,y))$
is not in $\nit{UJCQ}$, because the
bound and repeated variable
$y$ is for the changeable
attribute $B$. However, the
query
$\mathcal{Q}_2(y)\!: \ \ex x \ex z (R(x,y)\w R(x,z))$
is in $\nit{UJCQ}$: the only bound,
repeated variable is $x$ which is for the unchangeable attribute $A$. If variables
$x$ and $y$ are swapped in the first atom of $\mc{Q}_2$, the query is not \nit{UJCQ}.
\boxtheorem
\end{example}
We will use the $\nit{Count}(R)$ operator in queries \cite{Abiteboul}. It
returns the number of tuples in a relation $R$,
and will be applied to sets of tuples of the form $\{\bar x~|~C\}$, where
$\bar x$ is a tuple of variables,
and $C$ is a condition
involving a set of free variables that include those in
$\bar x$. More precisely, for an instance $D$, $\nit{Count}(\{\bar x~|~C\})$
takes on $D$ the numerical value $|\{\bar{c}~|~ D \models C[\bar{c}]\}|$. The variables
in $C$ that do not appear in $\bar{x}$ are intended to be existentially quantified.
A condition $C$ can be seen as a predicate defined by means of a Datalog query
with the $\neq$ built-in.
For motivation and illustration, we now present a simple example of rewriting using $\nit{Count}$.
Throughout the rest of this section, we use the notation of
Example \ref{ex:run} for the arguments of $\nit{TA}$.

\begin{example}  \label{ex:count} Consider $R[A,B]$, \
$m\!\!: \ R[A]\approx R[A]\ra R[B]\doteq R[B]$, and the \nit{UJCQ}
query $\mathcal{Q}(x,y,z)\!: \ R(x,y,z)$.
These are the extensions for $R$ and its (single) MRI:

\vspace{-1mm}
\begin{center}
\begin{tabular}{l|c|c|c|} \hline
$R$&$A$ & $B$ & $C$  \\ \hline
&$a_1$ & $b_1$ & $c_1$  \\
&$a_1$ & $b_2$ & $c_2$  \\
&$a_1$ & $b_2$ & $c_3$\\ \cline{2-4}
\end{tabular}
~~~~~~
\begin{tabular}{l|c|c|c|} \hline
$\mbox{MRI}$&$A$ & $B$ & $C$  \\ \hline
&$a_1$ & $b_2$ & $c_1$  \\
&$a_1$ & $b_2$ & $c_2$  \\
&$a_1$ & $b_2$ & $c_3$\\ \cline{2-4}
\end{tabular}
\end{center}

\noindent The set of resolved
answers to $\mathcal{Q}$ is
$\{(a_1,b_2,c_1),(a_1,b_2,c_2),$
$(a_1,b_2,c_3)\}$. The following query, directly posed to the (actually, any) initial instance,
returns the resolved answers. In it, $\nit{TA}$ stands for $\nit{TA}^{\{m\},(\cdot)}$. \vspace{-2mm}
\be
\hspace*{-3.8cm}\mathcal{Q}'(x,y,z)\!: \ex y'R(x,y',z) \ \wedge \ \fa y''[ \vspace{-1mm} \label{eq:rewrite}
\ee
$\nit{Count}\{
(x',y,z')~|~\nit{TA}(x,y',z,B,x',y,z',B) \wedge R(x',y,z')\} >$\\
$\nit{Count}\{(x',y'',z')~|~\nit{TA}(x,y',z,B,x',y'',z',B)\wedge R(x',y'',z')$\\
\phantom{poto}\hfill $\wedge y''\neq y\}].$

\vspace{1mm}\noindent As we saw in Section \ref{subsec:cycle}, the $\nit{TA}$ here can be specified by means of a Datalog query. Actually, the whole query can  be easily
expressed by means of a single Datalog query with aggregation\footnote{Count queries with group-by in Datalog can be
expressed by rules of the form $Q(\bar{x},\nit{count}(z)) \leftarrow B(\bar{x}')$, where $\bar{x} \cup \{z\} \subseteq
\bar{x}'$, $z \notin \bar{x}$, and $B$ is a conjunction of atoms.} and comparison as a built-in.

Intuitively, the first conjunct requires the
existence of a tuple $t$ with the same values as the answer for attributes $A$ and
$C$. Since
the values of these attributes are not
changed when going from the original instance
to an MRI, such a tuple must exist. However,
the tuple is not required to have the same
$B$ attribute value as the answer tuple,
because this attribute can be modified.
For example, $(a_1,b_2,c_1)$ is a resolved
answer, but is not in $R$. What makes it a
resolved answer is the fact that it is in
an equivalence class of value positions (consisting of
all three positions in the $B$ column of the instance) for which $b_2$
occurs more frequently
than any other value. This counting condition
on resolved answers is expressed by the second conjunct.
Attribute $B$ is the only changeable attribute, so it
is the only attribute argument to $\nit{TA}$, which specifies
the values to be merged.
Query (\ref{eq:rewrite}) can be computed in
polynomial time on any instance.
\boxtheorem
\end{example}
The  {\em Rewrite} algorithm in Table \ref{tab:rewrite}
uses a binary relation on attributes, that we now introduce.

\begin{definition}\label{def:atclosure}
Let $M$ be a set of MDs.
(a) The symmetric binary relation $\doteq_r$ is defined on attributes, as follows:
\ $R[A] \doteq_r S[B]$ iff
 there is $m \in M$ with  $R[A]\doteq S[B]$ appearing
on  the RHS of $m$'s arrow.\\
(b) $E_{R[A]}$ denotes the equivalence class of the reflexive,
transitive, closure of $\doteq_r$ that contains $R[A]$.
\boxtheorem
\end{definition}
\begin{example}\label{ex:atclosure} Let $M$ be the
set of MDs \vspace{-2mm}
\bea
&&R[A]\approx_1 S[B]\ra R[C]\doteq S[D],\nn\\
&&S[E]\approx_2 T[F] \w S[G]\approx T[H]\ra
S[D,K]\doteq T[J,L],\nn\\
&&T[F]\approx_3 T[H]\ra T[L,N]\doteq T[M,P].\nn
\eea
The equivalence classes of $T_{at}$
are $E_{R[C]} = \{R[C],S[D],T[J]\}$,
$E_{S[K]} = \{S[K],$ $T[L],T[M]\}$, and
$E_{T[N]} = \{T[N],T[P]\}$.
\boxtheorem
\end{example}
To emphasize the association between a variable
and a particular attribute, we sometimes subscript
the variable name with the name of the attribute. For example,
given a relation $R$ with attributes $A$ and $B$ and atom
$R(x,y)$, we sometimes write $x$ as $x_A$.
To express substitutions of
variables within lists of variables, we give the name of
the variable list, followed by the substitution in square
brackets. For example, the list of variables obtained from
the list $\bar v$ by substitution of variables from a subset
$S$ of the variables in $\bar v$ with primed variables is expressed as
$\bar v[v\ra v'~|~v\in S]$.

{\small
\begin{table}[ht]
\centering
\begin{tabular}{|p{8cm}|}\hline
\textbf{Input}: A query in $\nit{UJCQ}$ and a NI or HSC set of MDs
$M = \{m_1,...m_p\}$.\\
\textbf{Output}: The rewritten query $\mathcal{Q}'$.\\
1)~\quad Let $\mathcal{Q}(\bar t) : \ex \bar u \w_{1\leq i\leq n}R_i(\bar v_i)$
be the query.\\
2)~\quad Let $\nit{TA}$ denote $\nit{TA}^{M,(\cdot)}$\\
3)~\quad \textbf{For} each $R_i(\bar v_i)$\\
4)~\quad\qquad Let $C$ be the set of changeable attributes of $R_i$\\
~~~~\quad\qquad corresponding to a free variable in $\bar v_i$\\
5)~\quad\qquad \textbf{If} $C$ is empty\\
6)~\quad\qquad\qquad $Q_i(\bar v_i)\lar R_i(\bar v_i)$\\
7)~\quad\qquad \textbf{Else} \\
8)~\quad\qquad\qquad $\bar v_i'\lar \bar v_i[v_{iA}\ra v_{iA}'~|~A\in C]$\\
9)~\quad\qquad\qquad Let $\bar v_{iC}$ be the list of variables $v_{iA}$, $A\in C$\\
10)\quad\qquad\qquad $\bar v_{iC}'\lar \bar v_{iC}[v_{iA}\ra v_{iA}'~|~A\in C]$\\
11)\quad\qquad\qquad \textbf{For} each variable $v_{iA}$ in $\bar v_{iC}$\\
12)\quad\qquad\qquad\qquad \bf For \rm each attribute $R_j[B_k]\in E_A$\\
13)\quad\qquad\qquad\qquad\qquad Generate atom $R_j(\bar u_{jk}')$, with\\
~~~~\quad\qquad\qquad\qquad\qquad $\bar u_{jk}'$ a list of new variables\\
14)\quad\qquad\qquad\qquad\qquad $\bar u_{jk}\lar \bar u_{jk}'
[u_{jkR_j[B_k]}\ra v_{iA}]$\\
15)\quad\qquad\qquad\qquad\qquad $\bar w_{jk}\lar\bar
u_{jk}'[u_{jkR_j[B_k]}\ra v_{iA}'']$\\
16)\quad\qquad\qquad\qquad\qquad $C_{jk}^{A1}\lar
\nit{Count}\{\bar u_{jk}~|~\nit{TA}(\bar v_i',$\\
~~~~\quad\qquad\qquad\qquad\qquad $\bar R_i[A],u_{jk},R_j[B_k])\w R_j(\bar u_{jk})\}$\\
17)\quad\qquad\qquad\qquad\qquad $C_{jk}^{A2}\lar
\nit{Count}\{\bar w_{jk}~|~\nit{TA}(\bar v_i',$\\
~~~~\quad\qquad\qquad\qquad\qquad $R_i[A],\bar w_{jk},R_j[B_k])\w R_j(\bar w_{jk})$\\
~~~~\quad\qquad\qquad\qquad\qquad $\w v_{iA}''\neq v_{iA}\}$\\
18)\quad\qquad\qquad $Q_i(\bar v_i)\lar \ex \bar v_{iC}'\{R_i(\bar v_i')\w_{A\in C}\fa v_{iA}''
[\S_{j,k} C_{jk}^{A1}$\\
~~~~\quad\qquad\qquad $> \S_{j,k} C_{jk}^{A2}]\}$\\
19)\quad $\mathcal{Q}'(\bar t)\lar \ex \bar u\w_{1\leq i\leq n}Q_i(\bar v_i)$\\
20)\quad \textbf{return} $\mathcal{Q}'$\\\hline
\end{tabular}
\caption{{\em Rewrite} Algorithm}\label{tab:rewrite}
\end{table}
}

\vspace{-2mm}
\noindent {\em Rewrite}
outputs a rewritten query $\mc{Q}'$ for
an input consisting of a query $\mc{Q} \in \nit{UJCQ}$ and set of
NI or HSC MDs.
It rewrites the query by separately rewriting each conjunct $R_i(\bar v_i)$
in $\mc{Q}$.
If $R_i(\bar v_i)$ contains no free variables, then
it is unchanged (line 6). Otherwise, it is replaced with
a conjunction involving the same atom and additional
conjuncts which use the $\nit{Count}$ operator. The conjuncts
involving \nit{Count} express the condition that, for each
changeable attribute value returned by the query, this value
is more numerous than any other value in the same set of values
that is equated by the MDs. The \nit{Count} expressions contain new
local variables as well as a new universally quantified variable $v_{iA}''$.

\begin{example}\label{ex:alg}
We illustrate the algorithm with predicates $R[ABC], S[EFG], U[HI]$, the $\nit{UJCQ}$ query

\centerline{$\mathcal{Q}(x,y,z)\!: \ex t~u~p~q$ $(R(x,y,z)\w S(t,u,z)\w U(p,q))$;}

\noindent and the
NI MDs:
$R[A]\approx S[E]\ra R[B]\doteq S[F]$, and  $S[E]\approx U[H]\ra S[F]\doteq U[I]$.

Since the $S$ and $U$ atoms have no free variables
holding the values of changeable attributes, these
conjuncts remain unchanged (line 6).
The only free variable holding the value of a
changeable attribute is $y$. Therefore, line 8 sets
$\bar v_1'$ to $(x,y',z)$. Variable $y$ contains the value
of attribute $R[B]$.
The equivalence class $E_{R[B]}$ is
$\{R[B],S[F],U[I]\}$, so the loop at line 12
generates the atoms $R(x',y,z')$, $R(x',y'',z')$,
$S(t',y,z')$, $S(t',y'',z')$, $U(p',y)$, $U(p',y'')$.
The rewritten query is obtained by replacing in $\mc{Q}$ the
conjunct $R(x,y,z)$ by \ $\ex y' (R(x,y',z) \ \w \fa y''[$\\
$\nit{Count}\{(x',y,z')~|~\nit{TA}(x,y',z,R[B],x',y,z',R[B]) \ \w$\\
$ R(x',y,z')\} \ + \nit{Count}\{(t',y,z')~|~\nit{TA}(x,y',z,R[B],$\\
$ t',y,z',S[F])\w S(t',y,z')\}  + \nit{Count}\{(p',y)~|~\nit{TA}(x,y',z,$\\
$ R[B],p',y,U[I])\w U(p',y)\}  \ \ > \ \ \nit{Count}\{(x',y'',z')~| $\\
$ \nit{TA}(x,y',z,R[B],x',y'',z',R[B])\w R(x',y'',z')\w y''\neq y\} $\\
$  + \ \nit{Count}\{(t',y'',z')~|~\nit{TA}(x,y',z,R[B],t',y'',z',S[F]) \ \w$\\
$ S(t',y'',z')\w y''\neq y\} +\nit{Count}\{(p',y'')~|~\nit{TA}(x,y',z,R[B],$\\
$ p',y'',U[I])\w U(p',y'')\w y''\neq y\}]$. \boxtheorem
\end{example}
Notice that the resulting query in Example \ref{ex:alg}, and this is a general fact with the algorithm, can be easily translated into a Datalog query with the aggregate $\nit{Count}$
plus the built-ins $\neq$ and $>, +$, the last two applied to natural numbers resulting from counting. The FO part
can be transformed by means of the {\em Lloyd-Topor transformation} \cite{lloyd87}.

\begin{theorem}\label{thm:rewrite}
For a NI or HSC set of MDs $M$ and a $\nit{UJCQ}$ query
$\mathcal{Q}$, the
query $\mc{Q}'$ computed by the {\em Rewrite} algorithm
is efficiently evaluable and returns the resolved answers to $\mc{Q}$.
\boxtheorem
\end{theorem}
The rewriting algorithm does not depend on the dirty instance
at hand, but only on the MDs and the input query, and runs in polynomial time in the size of
$\mc{Q}$ and $M$.

In the next section, we will relate
$\nit{RA}_{\mathcal{Q},M}$ to {\em consistent query
answering} (CQA) \cite{B2006,bertossi11}. This connection and some known results in CQA
 will allow us to identify further tractable cases,
but also to establish the
intractability of $RA_{\mathcal{Q},M}$ for certain classes of queries and MDs.
The latter result implies that the tractability
results in this section cannot be extended to all
conjunctive queries.

\section{A CQA Connection}\label{subsec:NI2}

MDs can be seen as a new form of integrity constraint (IC), with a dynamic semantics.
An instance $D$ violates an MD $m$ if there are
unresolved duplicates, i.e.  tuples
$t_1$ and $t_2$ in $D$ that satisfy the similarity
conditions of $m$, but differ in value on some pairs of attributes
that are expected to be matched according to $m$. The instances that are
consistent with a set of MDs $M$ (or {\em self-consistent} from the point of view of the dynamic semantics) are
resolved instances of themselves with respect to $M$. Among classical
ICs, the closest analogues of MDs are functional dependencies (FDs).

Now, given a database instance $D$ and a set of ICs $\Sigma$, possibly not satisfied by $D$, {\em consistent query answering} (CQA) is the problem of characterizing
and computing the answers to queries $\mc{Q}$ that are true in all
{\em repairs} of $D$, i.e. the instances $D'$ that are consistent with $\Sigma$ and minimally differ
from $D$ \cite{Arenas99}.
 Minimal difference between instances can be defined in different ways. Most of the research in
CQA has concentrated on the case of the set-theoretic symmetric difference of instances, as sets of tuples,
which in the case of repairs is made minimal under set inclusion, as originally introduced in  \cite{Arenas99}.
Also the minimization of the  {\em cardinality} of this
set-difference has been investigated \cite{lopat07,kol09}.  Other forms of minimization measure the differences in
terms of changes of attribute values between $D$ and $D'$ (as opposed to entire tuples) \cite{fran01,wijsen05,fles10,berIS08}, e.g. the number of attribute updates can be used for comparison. Cf. \cite{B2006,chom07,bertossi11} for CQA.

Because of their practical
importance, much work on CQA has been done
for the case where  $\Sigma$ is a set of functional dependencies (FDs), and in particular
for sets, $\mc{K}$, of  key constraints (KCs) \cite{Chomicki05,fux07,Wijsen07,Wijsen09,Wijsen10},
with the distance being the set-theoretic
symmetric difference under set inclusion.
In this case, on which we concentrate in the rest of this section, a {\em repair} $D'$ of an instance $D$ becomes a
maximal subset of $D$ that satisfies
$\mc{K}$, i.e.  $D' \subseteq D, \ D' \models \mc{K}$, and there is no $D''$ with
 $D' \subsetneqq D'' \subseteq D$, with $D'' \models \mc{K}$ \cite{Chomicki05}.

 Accordingly, for a FO query $\mc{Q}(\bar{x})$ and a set of KCs $\mc{K}$, $\bar{a}$ is a {\em consistent answer} from $D$ to $\mc{Q}(\bar{x})$ wrt $\mc{K}$
 when $D' \models \mc{Q}[\bar{a}]$, for every repair $D'$ of $D$. For fixed $\mc{Q}(\bar{x})$ and $\mc{K}$,
  the {\em consistent query answering problem}
 is about deciding
membership in the set
$\nit{CQA}_{\mathcal{Q},\mc{K}} = \{(D,\bar{a})~|~\bar{a}\hbox{ is a consistent answer from}$ $D \mbox{ to }
\mathcal{Q}\hb{ wrt } \mc{K}\}.$ 

Notice that this notion of minimality involved in repairs wrt FDs  is tuple and set-inclusion
oriented, whereas the one that is implicitly related to MDs and MRIs via the matchings (cf. Definition \ref{def:minim})
is attribute and cardinality oriented.\footnote{Cf. \cite{jaf11} for a discussion of the differences between FDs and MDs seen as ICs, and their repair processes.}  However,  the connection can still be established. In particular, the following result can be obtained through a reduction and a result in
\cite[Thm. 3.3]{Chomicki05}.

\begin{theorem}\label{thm:Chom}
Consider the relational predicate $R[A,B,C]$, the MD \
$m\!: \ R[A] = R[A]\ra R[B,C]\doteq R[B,C]$,
and the non-\!$\nit{UJCQ}$ query $\mathcal{Q}\!:
\ex x \ex y \ex y' \ex z(R(x,y,c)\w R(z,y',d)\w y = y')$.
$RA_{\mathcal{Q},\{m\}}$ is $\nit{coNP}$-complete.\footnote{This result appeals to {\em many-one} or {\em Karp's reductions}, in contrast  to the {\em Turing reductions} used in Section \ref{subsec:com}.} \boxtheorem
\end{theorem}
For certain classes of conjunctive queries and
ICs consisting  of a single KC per relation, {\it CQA} is tractable.
This is the case for the  $\mathcal{C}_{\!\nit{forest}}$ class
of conjunctive queries
 \cite{fux07}, for which there is a FO
 rewriting methodology for computing
the consistent answers.
$\mathcal{C}_{\!\nit{forest}}$ excludes repeated relations (self-joins), and allows joins only
between non-key and key attributes.
Similar results were subsequently proved for a larger
class of queries that includes some queries with
repeated relations and joins between non-key
attributes \cite{Wijsen07,Wijsen09,Wijsen10}. The following result allows us to take advantage of tractability results for CQA in our
MD setting.

\begin{proposition}\label{thm:red}
Let $D$ be a database instance for
a single predicate $R$ whose set of attributes
is $\bar A \cup \bar B$, with  $\bar A \cap \bar B = \emptyset$; and $m$ the MD \
$R[\bar A] = R[\bar A]\ra R[\bar B]\doteq R[\bar B]$. \
There is a polynomial time
reduction from $\nit{RA}_{\mathcal{Q},\{m\}}$ to $\nit{CQA}_{\mathcal{Q},\{\kappa\}}$,
where $\kappa$ is the key constraint $\bar A\ra \bar B$. \boxtheorem
\end{proposition}
Proposition \ref{thm:red} can be easily
generalized to several relations with
one such MD defined on each. The reduction takes an instance $D$ for $\nit{RA}_{\mathcal{Q},\{m\}}$ and produces an instance $D'$ for
$\nit{CQA}_{\mathcal{Q},\{\kappa\}}$. The schema of $D'$ is the same as for $D$, but the extension of
the relation is changed wrt $D$ via counting. Definitions for those aggregations can be  inserted
into
query $\mc{Q}$, producing a rewriting $\nit{Q}'$. Thus, we obtain:

\begin{theorem} \label{theo:comp}Let $\mathcal{S}$
be a schema with $\mc{R} = \{R_1[\bar A_1,\bar B_1], \ldots,$ $ R_n[\bar A_n,\bar B_n]\}$ and
$\mc{K}$ the set of KCs \
$\kappa_i\!: \ R_i[\bar A_i]\ra R_i[\bar B_i]$.
Let $\mathcal{Q}$ be a FO query for which there is
a polynomial-time computable FO rewriting $\mathcal{Q}'$ for
computing the consistent answers to $\mathcal{Q}$.
 Then there is a polynomial-time computable
FO query $\mathcal{Q}''$ extended with aggregation\footnote{This is a proper extension of FO
query languages \cite[Chapter 8]{libkin}.} for
computing the resolved answers to $\mathcal{Q}$ from $D$ wrt the
set of MDs \
$m_i\!: \ R_i[\bar A_i] = R_i[\bar A_i] \ra R_i[\bar B_i]\doteq R_i[\bar B_i]$.
\boxtheorem
\end{theorem}
The aggregation in $\mathcal{Q}''$ in Theorem \ref{theo:comp}
arises from the {\em generic} transformation of the instance that
is used in the reduction involved in Proposition \ref{thm:red}, but here becomes implicit in the query.

We emphasize  that
$\mathcal{Q}''$ is {\em not} obtained using algorithm {\em Rewrite}
from Section \ref{sec:Tr}, which is not guaranteed to work
for queries outside the class $\nit{UJCQ}$. Rather,
a first-order transformation of the $R_i$ relations with
$\nit{Count}$ is composed with $\mathcal{Q}'$ to produce
$\mathcal{Q}''$. Similar to the {\em Rewrite} algorithm of Section \ref{sec:Tr}, it is used to
capture the most frequently occurring values for the changeable attributes for a
given set of tuples with identical values for the unchangeable attributes.

This theorem can be applied to decide/compute resolved answers in those cases where a FO rewriting for
CQA has been identified.  In consequence, it extends
the tractable cases identified in Section \ref{sec:Tr}. It can be applied
to queries that are not in $\nit{UJCQ}$.

\begin{example}
The query $\mathcal{Q}:~\ex x \exists y \exists z (R(x,y)\w S(y,z))$
is in the class $\mathcal{C}_{\!\nit{forest}}$
for relational predicates $R[A,B]$ and $S[C,E]$
and KCs $A\ra B$ and $C\ra E$.
By Theorem \ref{theo:comp} and the results in \cite{fux07},
 there is a polynomial-time computable
FO query with counting that returns the resolved answers
to $\mc{Q}$ wrt the MDs $R[A] = R[A]\ra R[B]\doteq R[B]$
and $S[C] = S[C]\ra S[E]\doteq S[E]$. Notice that
$\mathcal{Q}$ is not in $\nit{UJCQ}$, since the bound
variable $y$ is associated with the changeable attribute
$R[B]$. \boxtheorem
\end{example}

\section{Conclusions}\label{sec:con}

Matching dependencies specify both a set of integrity constraints that
need to be satisfied for a database to be free of unresolved duplicates,
and, implicity, also a procedure for resolving such duplicates. Minimally resolved instances \cite{jaf11}
define the end result of this duplicate resolution process.
In this paper we considered the problem of computing the answers to a query
that persist across all MRIs (the resolved answers). In particular, we
studied query rewriting methods for obtaining these answers from the
original instance containing unresolved duplicates.

Depending on syntactic criteria on MDs and queries,
trac-table and intractable cases of resolved query answering were identified. We discovered
the first dichotomy result in this area.
In some of the tractable cases, the original query can be rewritten into a new,
polynomial-time evaluable query that returns
the resolved answers when posed to the original instance.
It is interesting that the rewritings make use
of counting and recursion (for the transitive closure).
The original queries considered in this paper are all conjunctive.
Other classes of queries will be considered in future work.

We established interesting connections between resolved query answering wrt MDs and
consistent query answering. There are still many issues to explore in this direction, e.g. the
possible use
of logic programs with stable model semantics to specify the MRIs, as  with database repairs \cite{ArenasBC03,BarceloBB01,greco03}.

We have proposed some efficient algorithms for resolved query answering. Implementing them and
experimentation are also left for future work. Notice that those algorithms use different forms of transitive closure.
 To avoid unacceptably slow
query processing, it may be necessary to
compute  transitive closures off-line
and store them.
The use of Datalog with aggregation can be investigated in this direction.

In this paper we have not considered  matching attribute values,
whenever prescribed by the MDs, using matching
functions \cite{icdt11}. This element adds an entirely new dimension to the semantics and the problems investigated here.

\vspace{2mm}\noindent
{\bf Acknowledgements:} Research funded by NSERC Discovery,
and the BIN NSERC Strategic Network on Business Intelligence (project ADC05).
L. Bertossi is a Faculty Fellow of the IBM CAS.

\bibliographystyle{abbrv}

\appendix

\section{Auxiliary Results and Proofs} \label{sec:proofs}

\vspace{2mm}
For several of the proofs below, we need some auxiliary definitions and results.

\begin{lemma}\label{lem:tc}
Let $D$ be an instance and let
$m$ be the MD
\bea
R[\bar A]\approx S[\bar B]\ra R[\bar C]\doteq R[\bar E]\nn
\eea
 An instance $D'$ obtained by
changing modifiable attribute values of $D$
satisfies $(D,D')\vD m$ iff
for each equivalence
class of $T_m$, there is a constant vector
$\bar v$ such that,
for all tuples $t$ in the equivalence class,
\bea
t'[\bar C] = \bar v~~\hbox{if } t\in R(D)\nn\\
t'[\bar E] = \bar v~~\hbox{if } t\in S(D)\nn
\eea
where $t'$ is the tuple in $D'$ with the
same identifier as $t$.
\end{lemma}
{\em Proof}:Suppose $(D,D')\vD m$. By Definition \ref{def:new}, for each
pair of tuples $t_1\in R(D)$ and $t_2\in S(D)$ such that
$t_1[\bar A]\approx t_2[\bar B]$,
$$
t_1'[\bar C] = t_2'[\bar E]\nn
$$
Therefore, if $T^\approx(\bar t_1, \bar t_2)$ is true, then
$t_1'$ and $t_2'$ must be in the transitive closure
of the binary relation expressed by
$t_1'[\bar C] = t_2'[\bar E]$. But the
transitive closure of this relation is the
relation itself (because of the transitivity
of equality). Therefore, $t_1'[\bar C] = t_2'[\bar E]$.
The converse is trivial. \boxtheorem

We require the
following definitions and lemma.

\begin{definition}\label{def:coverset}
Let $S$ be a set and let $S_1$, $S_2$,...$S_n$ be
subsets of $S$ whose union is $S$. A {\em cover subset} is a subset
$S_i$, $1\leq i\leq n$, that is in a smallest
subset of $\{S_1,S_2,...S_n\}$ whose union is $S$.
The problem {\em Cover Subset (CS)} is the problem
of deciding, given a set $S$, a set of subsets
$\{S_1,S_2,...S_n\}$ of $S$, and an subset
$S_i$, $1\leq i\leq n$, whether or not $S_i$
is a cover subset.\boxtheorem
\end{definition}

\begin{lemma}\label{lem:minset}
CS and its complement are $\nit{NP}$-hard.
\end{lemma}
{\em Proof}:The proof is by Turing reduction from
the minimum set cover problem, which is $\nit{NP}$-complete.
Let $O$ be an oracle for CS.
Given an instance of minimum set cover consisting
of set $S$, subsets $S_1$, $S_2$,...$S_n$ of $S$,
and integer $k$, the following algorithm determines
whether or not there exists a cover of $S$ of size
$k$ or less. The algorithm queries $O$ on $(S,\{S_1,...S_n\},S_i)$
until a subset $S_i$ is found for which $O$ answers
yes. The
algorithm then invokes itself recursively on the
instance consisting of set $S\backslash S_i$,
subsets \\$\{S_1,...S_{i-1},S_{i+1},...S_n\}$, and
integer $k-1$. If the input set in a recursive call
is empty, the algorithm halts and returns yes, and
if the input integer is zero but the set is nonempty,
the algorithm halts and returns no. It can be shown
using induction on $k$ that this algorithm returns the
correct answer. This shows that CS is $\nit{NP}$-hard.
The complement of CS is hard by a similar proof, with the oracle for
CS replaced by  an oracle for the complement of CS.
\boxtheorem

\vspace{2mm}
\defproof{Lemma \ref{thm:filter}}{
We assume that an attribute of both $R$ and $S$
in $\nit{RHS}(m_1)$ occurs in $\nit{LHS}(m_2)$.
The other cases are similar.
For each L-component of $m_1$,
there is an attribute of $R$ and an attribute of $S$
from that L-component
in $\nit{LHS}(m_2)$. Let $t_1\in R$ be a
tuple not in a singleton equivalence class of $T_{m_1}$.
Suppose there exist two conjuncts in $\nit{LHS}(m_1)$ of the
form $A\approx B$ and $C\approx B$. Then it must hold that
there exists $t_2\in S$ such that
$t_1[A]\approx t_2[B]$ and $t_1[C]\approx t_2[B]$ and by
transitivity, $t_1[A]\approx t_1[C]$. More generally, it follows
from induction that $t_1[A]\approx t_1[E]$ for
any pair of attributes $A$ and $E$ of $R$ in the same L-component of $m_1$.

We now prove that for any pair of tuples $t_1,t_2\in R$
satisfying $T_{m_2}(t_1,t_2)$ such that each of
$t_1$ and $t_2$ is in a non-singleton equivalence class
of $T_{m_1}$, for any instance $D$ it
holds that $T_{m_1}(t_1,t_2)$. By symmetry, the same result
holds with $R$ replaced with $S$.
Suppose for a contradiction that $T_{m_2}(t_1,t_2)$ but
$\neg T_{m_1}(t_1,t_2)$ in $D$.
 Then it must be true that
$t_1[\bar A]\not\approx t_2[\bar A]$, since, by assumption,
there exists a $t_3\in S$ such that
$t_1[\bar A]\approx t_3[\bar B]$, which together with $t_1[\bar A]\approx t_2[\bar A]$
would imply $T_{m_1}(t_1,t_2)$. Therefore, there must be an
attribute $A'\in \bar A$ such that $t_1[A']\not\approx t_2[A']$,
and by the previous paragraph and transitivity, $t_1[A'']\not\approx t_2[A'']$ for
all $A''$ in the same L-component of $m_1$ as $A'$. By
transitivity of $\approx_2$, this implies $\neg T_{m_2}(t_1,t_2)$,
a contradiction.

A resolved instance is obtained in two updates.
Let $T_{m_2}^0$ and $T_{m_2}^1$ denote $T_{m_2}$
before and after the first update, respectively.
The first update involves setting the attributes in $\nit{RHS}(m_1)$
to a common value for each non-singleton equivalence class of $T_{m_1}$.
The relation $T_{m_2}^1$ will depend on these common
values, because of accidental
similarities. However, because of the property
proved in the previous paragraph, this dependence
is restricted. Specifically, for each equivalence
class $E$ of $T_{m_2}^1$, there is at most
one non-singleton equivalence class $E_1$ of $T_{m_1}$
such that $E$ contains tuples of $E_1\bc R$ and
at most one non-singleton equivalence class $E_2$ of $T_{m_1}$
such that $E$ contains tuples of $E_1\bc S$.
 A given choice of update values
for the first update will result in a set of sets of tuples
from non-singleton equivalence classes of $T_{m_1}$ (ns tuples)
that are equivalent under $T_{m_2}^1$.
Let $K$ be the set of all such sets of
ESs. Clearly, $|K|\in O(n^2)$,
where $n$ is the size of the instance.

Generally, when the instance is updated according to
$m_1$, there will be more than one set of choices of
update values that will lead to the ns tuples being
partitioned according to a given $k\in K$. This is
because an equivalence class of $T_{m_2}^1$ will also
contain tuples in singleton equivalence classes of $T_{m_1}$
(s tuples), and the set of such tuples contained in
the equivalence class will depend on the update values
chosen for the modifiable attribute values in the ns tuples
in the equivalence class.
For a set $E\in k$, let $E'$ denote
the union over all sets of update values for $E$ of the equivalence classes
of $T_{m_2}^1$ that contain $E$ that result from choosing that set of
update values. By transitivity and
the result of the second paragraph, these $E'$ cannot overlap
for different $E\in k$.
Therefore, minimization of the change produced by the two
updates can be accomplished by minimizing the change for
each $E'$ separately. Specifically, for each equivalence class
$E$, consider the possible sets of update values for the attributes in
$\nit{RHS}(m_1)$ for tuples in $E$. Call two such sets of values equivalent if they
result in the same equivalence class $E_1$ of $T_{m_2}^1$.
Clearly, there are at most $O(n^c)$ such sets of ESs of values,
where $c$ is the number of R-components of $m_1$.
Let $V$ be a set consisting of one set of values $v$ from each set of sets of equivalent values.
For each set of values $v\in V$, the minimum number of changes produced
by that choice of value can be determined as follows.
The second application of $m_1$ and $m_2$ updates to a common
value each element in a set $S_2$ of sets of value positions that can be
determined using
lemma \ref{lem:tc}. The update values that result in
minimal change are easy to determine. Let $S_1$ denote the corresponding set of sets of value positions
for the first update. Since the second update ``overwrites" the first,
the net effect of the first update is to change to a common value the value
 positions in
each set in $\{S_i~|~S_i = S\backslash \bigcup_{S'\in S_2} S',~S\in S_1\}$.
It is straightforward to determine the update values that yield
minimal change for each of these sets.
This yields the minimum number of changes for this choice of $v$. Choosing
$v$ for each $E$ so as to minimize the number of changes allows the
minimum number of changes for resolved instances in which
the ns tuples are partitioned according to $k$ to be determined
in $O(n^c)$ time. Repeating
this process for all other $k\in K$ allows the determination
of the update values that yield an MRI in $O(n^{c+2})$ time.
Since the values to which each value in the instance can change
in an MRI can be determined in polynomial time, the result
follows.}

\vspace{2mm}
\defproof{Theorem \ref{thm:conv}}{
For simplicity of the presentation,
we make the assumption that the domain of all
attributes is the same. All pairs of
distinct values in an instance are dissimilar.
Wlog, we will assume that part (a) of
Theorem \ref{thm:combine} does not hold.
Let $E$ and $L$ denote an ES and
an L-component that violate
part (a) of Theorem \ref{thm:combine}.
 We prove the theorem separately for the following
three cases: (1) There exists such an $E$ that contains only attributes of
$m_1$, (2) there exists such an
$E$ that contains both attributes not
in $m_1$ and attributes in $m_1$, and (3) (1) and (2) don't hold
(so there exists such an $E$ that contains only attributes
not in $m_1$). Case (1) is divided
into two subcases: (1)(a) Only one
R-component of $m_1$ contains attributes of
$E$ and (1)(b) more than one R-component
contains attributes of $E$.

Case (1)(a):  We reduce an instance of the compliment of CS
(cf. definition \ref{def:coverset})
to this case, which is $\nit{NP}$-hard by lemma \ref{lem:minset}.
Let $F$ be an instance of CS with
set of elements $U = \{e_1,e_2,...e_n\}$ and set of
subsets $V = \{f_1,f_2,...f_m\}$. Wlog, we assume
in all cases
that each element is contained in at least two sets.
 With each subset in $V$ we
associate a value in the set $K = \{k_1,k_2,...k_m\}$.
With each element in $U$ we associate a value in the
set $P = \{v_1,v_2,...v_n\}$. The instance will also
contain values $b$ and $c$.

Relations $R$ and $S$ each contain a set $S_i$ of tuples for each
$e_i$, $1\leq i\leq n$. Specifically, there is a tuple in
$S_i$ for each value in $K$ corresponding to a set to which
$e_i$ belongs.
On attributes in $L$, all tuples in $S_i$
take the value $v_i$. There is one
tuple for each value in $K$ corresponding to a set to
which $e_i$ belongs that has that value as the value
of all attributes in the R-component
of $m_1$ that contains an attribute in $E$.
On all other attributes, all tuples in all $S_i$
take the value $b$.

Relation $S$ also contains a set $G_1$ of $m$
other tuples. For each value in $K$, there is a
tuple in $G_1$ that takes this value on all
attributes $A$ such that there
is an attribute $B\in E$ such that $B\approx A$
occurs in $m_2$. This tuple also takes this
value on some attribute $Z$ of $S$ in
$\nit{RHS}(m_2)$. For all other attributes,
all tuples in $G_1$ take the value
$b$.

A resolved instance is obtained in two updates.
We first describe a sequence of updates that
will lead to an MRI.
It is easy to verify that the equivalence classes
of $T_{m_1}$ are the sets $S_i$.
In the first
update, the effect of applying $m_1$ is to update
all modifiable values of attributes in $\nit{RHS}(m_1)$
within each equivalence class,
which are values
of attributes within the R-component of
$m_1$ that contains an attribute of $E$,
to a common value.
For some minimum set cover $C$, we choose
as the update value for a given $S_i$
the value associated with a set in $C$
containing $e_i$.

Before the
first update, there is one equivalence class of
$T_{m_2}$ for each value in $K$.
Let $E_k$ be the equivalence class for the value
$k\in K$.
$E_k$ contains all the
tuples in $R$ with $k$ as the value for the attributes
in $E$, as well as a tuple in $G_1$ with $k$ as
the value for $Z$.
The only R-component of $m_2$ the values of whose attributes
are modifiable for tuples in $E_k$ is the one
containing the attribute $Z$. If $k$ is the
value in $K$ corresponding to a set in the
minimum set cover $C$, then we choose $b$ as the
common value for this R-component. Otherwise,
we choose $k$.

After the first update, applying $m_1$ has
no effect, since none of the values of attributes
in $\nit{RHS}(m_1)$ are modifiable. Each equivalence class
of $T_{m_2}$ consists of a set of sets $S_i$
and a tuple of $G_1$. Specifically, for each
update value that was chosen for the modifiable attributes
of $\nit{RHS}(m_1)$ in the first update there is
an equivalence class that includes the set of
all $S_i$ whose tuples' $\nit{RHS}(m_1)$ attributes were updated to that value
as well as the tuple of $G_1$ containing this
value. Given the choices of update values in the
previous update, it is easy to see that the
values of all attributes in $\nit{RHS}(m_2)$ for
tuples in these equivalence classes are modifiable
after the first update unless all the values are
$b$. We choose $b$ as the update
value.

It can easily be seen that, in this update process,
the changes made to values of attributes in $\nit{RHS}(m_2)$
in the first update are overwritten by those made
in the second update. Therefore, the total number
of changes made in the two updates is the number
$n_1$ of changes made to the values of attributes
in $m_1$ during the first update plus the number of changes $n_2$
made to the attributes of $m_2$ during the second update.
The only attributes of $m_2$ whose values change
to a value different from the original instance in the
second update are those of attribute $Z$ for
tuples in $G_1$. Since these
values change iff they occur within a tuple
containing one of the update values for the $S_i$,
 $n_2$ is the
size of a minimum set cover.

When $m_1$ is applied to the instance in the first
update, the set of values of attributes in the
R-component of $m_1$ that contains an attribute
of $E$ for each set of tuples $S_i$
is updated to a common value. Before this update,
each such set of values includes the values of
the sets to which $e_i$ belongs.
For an arbitrary first update of
the instance according to $m_1$,
consider the set $I$ of $S_i$ for which the
update value occurs within the set.
We claim that for an MRI the set of update
values for $I$ must correspond to a minimum
set cover for the set of all $e_i$ such that
$S_i\in I$. Indeed, if these values did not
correspond to a minimum cover set,
then an instance with fewer changes could
be obtained by choosing them to be a minimum
cover set. Furthermore, an update in which
$I$ does not include all $S_i$ cannot
produce a resolved instance with fewer changes
than our update process. This is because,
for each $S_i$ not in $I$, at least one
additional value from among the values of
attributes in $\nit{RHS}(m_1)$ for tuples in $S_i$
was changed relative to our update process.
Thus, the update could be changed so
that all $S_i$ are in $I$ without increasing
the number of changes, and the resulting update
would have at least as many changes as one
in which the set of update values corresponds to
a minimum set cover. This implies that
a value from $K$ occurs as a value of attribute
$Z$ in all MRIs iff the value does not
correspond to a cover set. Thus, RAP is
hard for the query $\pi_ZS$.

Case (1)(b): Let $F$ be the min set cover instance
from case (1)(a), and define sets of values $K$
and $P$ as before. In addition, define a
set $Y$ of $2n$ values and
values $a$, $c$.

Relations $R$ and $S$ contain a set $S_i$ for each $e_i$, $1\leq i\leq n$
as before. However, these sets now contain one
more tuple than in case (1)(a). On attributes
in $L$ tuples in each $S_i$ take the same value as in
case (1)(a).  Let $\{k_1',k_2',...k_{|S_i|}'\}$
and $\{k_1'',k_2'',...k_{|S_i|}''\}$ be lists of
all the values in $K$ corresponding to sets to
which $e_i$ belongs such that $k_i' = k_{i\bmod{|S_i|}+1}''$.
For some R-component of $m_1$ containing an
attribute of $E$, for each $1\leq j\leq |S_i|$,
there is a tuple in $S_i$ that takes the value $k_j'$ on
all attributes in this component and the value $k_j''$ on
all attributes of all other R-components of $m_1$
containing attributes of $E$. (We do this to ensure
that all tuples in all $S_i$ are in singleton equivalence
classes of $T_{m_2}$ before the first update, and so
their values are not updated by the application of
$m_2$ in this update.)
There is also a tuple that takes the value $a$ on
all attributes of all R-components of $m_1$ containing
attributes of $E$. On all other attributes, all tuples in all $S_i$
take the value $b$.

Relation $R$ also contains a set $G_1$ of $2n$ other tuples.
For each value
in $Y$, there is a tuple in $G_1$ with that
value as the value of all attributes of $R$ in $L$.
 There are $2n$ tuples
with value $a$ for all attributes in $E$.
For all attributes of $R$ in $\nit{RHS}(m_2)$,
all tuples in $G_1$ take the value $c$. On all
other attributes, tuples in $G_1$ take the value
$b$.

Relation $S$ also contains a set $G_2$ of $m+1$
other tuples. For each value in $K$, there is a
tuple in $G_2$ that takes this value on all
attributes $A$ such that there
is an attribute $B\in E$ such that $B\approx A$
occurs in $m_2$. This tuple also takes this
value on some attribute $Z$ of $S$ in
$\nit{RHS}(m_2)$. There is also a tuple $t_1$
 which takes the value $a$ on all attributes
$A$ such that there
is an attribute $B\in E$ such that $B\approx A$
occurs in $m_2$, and the value $c$ on $Z$.
 For all other attributes,
all tuples in $G_2$ take the value
$b$ except $t_1$ which takes the
value $c$.

As in case (1)(a), a resolved instance is obtained in two updates.
We now describe a series of updates that leads to an MRI.
The equivalence classes of $T_{m_1}$ are the sets $S_i$
as before. The sets of modifiable values in $\nit{RHS}(m_1)$
are the sets of values of tuples in $S_i$ for attributes in an
R-component of $m_1$ that contains an
attribute of $E$. We again choose the update values
to correspond to a minimum set cover, and we choose
the same update value for all R-components for a given
$S_i$.

Before the first update, there is one
equivalence class of $T_{m_2}$ containing
all tuples that have value $a$ for attributes
in $E$.
The values of all attributes in $\nit{RHS}(m_2)$
are modifiable for tuples in this equivalence class.
We choose $c$ as the common value. After
the first update, the equivalence classes
of $T_{m_2}$ are as in case (1)(a), and
we choose the same update values as before.

As in case (1)(a),
the changes made to values of attributes in $\nit{RHS}(m_2)$
in the first update are overwritten by those made
in the second update. As in that case, this implies
that the total number of changes is the number of
changes made to the attributes of $m_1$ during the
first update plus the number of subsets in a
minimum set cover.

If the update value chosen for the $\nit{RHS}(m_2)$
attributes of the equivalence
class of $T_{m_2}$ in the first update is not $c$,
the resulting resolved instance cannot be
an MRI. Indeed, suppose that there is a different
value that can be used to obtain an MRI. If this value is
chosen, then the number of changes to
the values of attributes of $\nit{RHS}(m_2)$ for tuples
in $G_1$ resulting from the
update is at least $2n$. Since our update process
makes at most $n$ changes to these values
and the minimum number of changes to the values of attributes
of $\nit{RHS}(m_1)$, this implies that these values must
be modifiable after the first update so that
they can be changed back to their original value
in the second update. Modifiability
can only be achieved by updating the values of
attributes in $\nit{RHS}(m_1)$ to $a$ for some $S_i$ in
the first update. However, this would result in
at least 3 changes to values in tuples in $S_i$
in the second update, since these tuples would
then be in the same equivalence class of $T_{m_2}$
as the tuples in $G_1$. Because other choices
of update values for $S_i$ in the first update
result in only 1 change, this cannot produce an
MRI. In fact, this shows that, even if the first
update using $m_2$ is
kept the same as in our update process, using $a$
as the update value for the $\nit{RHS}(m_1)$
attributes of $S_i$ in the first update will not
produce an MRI.

When $m_1$ is applied to the instance
in the first update, the set of values
for the attributes in an R-component of $m_1$ for a given $S_i$
are updated to a common value. Suppose that
for each R-component, the update value is a value in $K$
that is in the set, and the update
values for the R-components are not all the
same. It is straightforward to show that
this implies that all the tuples in $S_i$
will be in singleton equivalence classes of
$T_{m_2}$ after the first update, and so will not be changed in the
second update. As we have shown, for any update
process leading to an MRI, at least one change
must be made to the values of attributes in $\nit{RHS}(m_2)$
for tuples in $S_i$ during the first update.
Since these changes are undone in our update
process, the number of updates to the tuples
in $S_i$ is at least one greater than in
our update process. The result now follows from
exactly the same argument used in case (1)(a),
except with the additional requirement for
$S_i$ in $I$ that their update values are the
same for all R-components of $m_1$.

Case (2): For simplicity of the presentation,
we will assume that there exists only one
attribute $A$ in $E$ not in $m_1$. Let $F$ be the min set cover instance
from case (1)(a), and define sets of values $K$
and $P$ as before. In addition, define $m$
sets $Y_i$, $1\leq i\leq m$, of $2n$ values and
values $a$, $b$, and $c$.

Relations $R$ and $S$ contain a set $S_i$ for each $e_i$,
$1\leq i\leq n$,
as before. However, $S_i$ now contains two
tuples for each set to which $e_i$ belongs.
On attributes
in $L$, tuples in each $S_i$ take the same value as in
case (1)(a). Let $K' = \{k_1',k_2',...k_{|S_i|}'\}$
and $K'' = \{k_1'',k_2'',...k_{|S_i|}''\}$ be lists as
defined in case (1)(b). For each value $k_i'\in K'$, there
are two tuples in $S_i$ that take this value on all
attributes in all R-components of $m_1$ containing
an attribute of $E$. On the attribute $A$, one of
these tuples takes the value $k_i'$ and the other
takes the value $k_i''$. On all other attributes,
all tuples in all $S_i$
take the value $b$.

Relation $R$ also contains a set $G_1$ of $4nm$ other tuples.
For each value
in each $Y_i$, $1\leq i\leq m$, there are two tuples $t_1$ and $t_2$
in $G_1$ with that
value as the value of all attributes of $R$ in $L$.
Tuple $t_1$ takes the value $a$ for
all attributes in $E$ except $A$, and $t_2$ takes
the value in $V$ corresponding to $S_i$ on these attributes.
For all attributes of $R$ in $\nit{RHS}(m_2)$,
$t_1$ takes the value $c$ and $t_2$ takes
the value in $V$ corresponding to $S_i$.
On attribute $A$, both tuples take the value
in $V$ that corresponds to $S_i$. On all
other attributes, tuples in $G_1$ take the value
$b$.

Relation $S$ also contains a set of tuples $G_2$
containing $2nm$ tuples.
For each value in each $Y_i$, $1\leq i\leq m$,
there is a tuple in $G_2$ that takes the value
on all attributes in $L$. On all attributes in
all R-components of $m_1$ that contain an attribute
of $E$, tuples in $G_1$ take the value $a$. For all
attributes of $S$ in $\nit{RHS}(m_2)$,
all tuples in $G_2$ take the value $c$.
On all other attributes, tuples in $G_1$ take the value
$b$.

Relation $S$ also contains a set of tuples $G_3$
containing $m$ tuples. For each value in $K$, there is a
tuple in $G_3$ that takes this value on all
attributes $A$ such that there
is an attribute $B\in E$ such that $B\approx A$
occurs in $m_2$. The tuple also takes this value
 on some attribute $Z$ of $S$ in
$\nit{RHS}(m_2)$. For all other attributes,
all tuples in $G_3$ take the value
$b$.

As in case (1), a resolved instance is obtained in two updates.
We now describe a series of updates that leads to an MRI.
The equivalence classes of $T_{m_1}$ are the sets $S_i$,
as well as $2nm$ sets of 3 tuples, two from $G_1$ and one
from $G_2$ that take the same value on attributes
in $L$. For the $S_i$, we choose the update
values for attributes in $\nit{RHS}(m_1)$ in the same
way as in case (1)(b). For the other equivalence
classes, we choose the update value $a$.

Before the first update, the only
equivalence classes of $T_{m_2}$ such that
the $\nit{RHS}(m_2)$ attribute values are modifiable
are those containing tuples from the sets $S_i$.
Each of these equivalence classes includes
tuples in $S_i$ that take a given value $v$ from
$V$ on all attributes in $E$ (including $A$),
as well as those tuples of $G_1$ that take the
value $v$ on these attributes and the tuple
from $G_3$ that contains this value.
Call such an equivalence class $E_v$. We choose $v$ as
the update value for each $E_v$.

After the first update, the equivalence classes
of $T_{m_2}$ are similar to those in case (1).
As in that case, we choose update values in
the second update so as to overwrite the
the changes made to values of attributes in $\nit{RHS}(m_2)$
in the first update. This implies
that the total number of changes is the number of
changes made to the attributes of $m_1$ during the
first update plus the number of subsets in a
minimum set cover.

We now show that, as in case (1), the
value in a tuple in $G_3$ that corresponds
to a given set in $V$ changes in some MRI iff that set is in
a min set cover. Consider the first update produced
by the application of $m_2$. Suppose that
the update value for an equivalence class $E_v$ is not $v$, and assume
for a contradiction that this leads to an MRI.
 This update would
result in at least $2n$ changes in the values
of tuples in $G_1$, and thus would produce
at least $n$
more changes than the
maximum number of changes that our update
process could produce. Therefore, at least some of
the values of
tuples in $G_1$ in this equivalence class
must be modifiable after the first update,
so that they can be restored to their original
values. This implies that, in the update produced by $m_1$,
 the update value chosen
for any such modifiable tuple
cannot be $a$, or it would be in a singleton equivalence
class of $T_{m_2}$ after the update.
However, not choosing $a$ as the update value would
result in at least one more change relative to
our update process. This is because the updated values
include at least one more $a$ than any other value.
Thus, the first update value for the equivalence
classes of $T_{m_2}$ must be chosen as in our update
process in order to obtain an MRI.

Consider the update resulting from the application
of $m_1$. If an update to an equivalence class
involving tuples of $G_1$ and $G_2$ does not use
the value $a$, then the resolved instance obtained
cannot be an MRI. This is because using any other
choice of value would result in at least one more change
in these tuples relative to our update process in the first update,
and cannot result in fewer updates in the second
update since choosing $a$ makes the values in tuples in the
equivalence class unmodifiable. The result now follows
from an argument similar to that of case (1).

Case (3): Let $F$ be the CS instance
from case (1)(a), and define sets of values $K$
and $P$ as before. Let $E'$ be an ES containing attributes of $m_1$. Since
the MDs are interacting, there must be at least
one such ES, and by assumption, it
must contain an attribute of $\nit{LHS}(m_1)$. Let $C_1$
denote some R-component of $m_1$ that contains
an attribute of $E'$, and let $p$ denote the number
of attributes in $C_1$. Let $C_2$ denote some R-component
of $m_2$. Let $q$ be
the number of attributes of $R$ in $C_2$. We define a set $W$
of values of size $p^2$, and $mn$
sets $Y_{ij}$, $1\leq i\leq m$, $1\leq j\leq n$, of $p+q$ elements
each. We also define a value $a$.

Relations $R$ and $S$ contain a set $S_i$ for each set $f_i$,
$1\leq i\leq m$, in $V$. For each element $e_j$ in $f_i$,
$S_i$ contains a set $S_{ij}$ of $p+q$ tuples. On all attributes
of $L$, all tuples in $S_i$ take the value $k_i$
in $K$ corresponding to $f_i$. For any given $S_{ij}$,
for a set of $p$ tuples in $S_{ij}$, each value
in $W$ occurs once as the value of an attribute
in $C_1$ for a tuple in the set. All other tuples
in $S_{ij}$ take the value $a$ on all attributes
in $C_1$. For each value
in $Y_{ij}$, there is a tuple in $S_{ij}$ that
takes the value on all attributes in $C_2$.
On all attributes of $E$, each tuple in $S_{ij}$,
$1\leq i\leq m$, takes the value $v_j$ in $P$ that
is associated with $e_j$. On all other attributes,
all tuples in $S_i$ take the value $a$.

Relation $S$ also contains a set of tuples $G_1$.
For each pair $(f_i,e_j)\in V\times U$, there
is a set of tuples $X_{ij}$ in $G_1$ of size $p+q$.
For all attributes of $S$ in the L-component containing
the attributes of $E$, each $X_{ij}$ takes the
value $v_j$ in $P$ associated with $e_j$. For
each value in $Y_{ij}$, there is a tuple in $X_{ij}$
that takes this value on all attributes of $C_2$.
On all other attributes, all tuples in $G_1$ take
the value $a$.

A resolved instance is obtained in two updates. The equivalence
classes of $T_{m_1}$ are the sets $S_i$. The effect of the
first update is to change all values of all attributes
in $C_1$ for tuples in $S_i$ to a common value. It
is easy to see that if the update value is not $a$, then
all tuples in $S_i$ will be in singleton equivalence
classes of $T_{m_2}$ after the update. Thus, the equivalence
classes of $T_{m_2}$ after the update are
$\bigcup_J S_{ij}$, $1\leq j\leq n$, where
$J \equiv\{i~|~a\hbox{ was chosen as the update value for }
S_i\}$.
If the update value $a$ is chosen for $S_i$ for some $i$,
 we say that $S_i$ is
{\em unblocked}. Otherwise, it is {\em blocked}.

Consider a blocked $S_i$.
In the first update, the minimum number of changes
to values for attributes in $\nit{RHS}(m_1)$ is $p(p+q)k - 1$,
where $k$ is the number of elements in $f_i$.
Minimal change of the values of attributes
in $C_2$ for tuples in an equivalence
class of $T_{m_2}$ is achieved by updating to one of the
original values. The number of changes to values of attributes
in $\nit{RHS}(m_2)$ for tuples in $S_i$ depends on the number of
sets $S_{ij}$ that are contained in $S_i$ that
contain the tuple with this update value. The greater this
number, the fewer the changes. We will take this into account
later, but we ignore it for now and assume that the values
of attributes of $\nit{RHS}(m_2)$ are updated to values
outside the active domain in the first update. Under this assumption,
the resulting upper bound
on the number of changes is
$q^2k + d(p+q)k$, where $d$ is the number of attributes
of $S$ in $C_2$. Since all
tuples in $S_i$ are in singleton equivalence classes of
$T_{m_2}$ after the first update, the second update produces
no further changes. Therefore, the number of changes
of values for tuples in $S_i$ is at most
$p(p+q)k - 1 + q^2k + d(p+q)k$.

For an unblocked $S_i$, the minimum number of changes
to values for attributes in $\nit{RHS}(m_1)$ is $p^2k$. Since the
second update ``overwrites" the  first, the number
of changes to the values of attributes in $\nit{RHS}(m_2)$ is
the number of changes produced in the second update.
Minimal change of the values of attributes
in $C_2$ for tuples in an equivalence
class of $T_{m_2}$ is achieved by updating to one of the
original values for these tuples and attributes.
A set $S_{ij}$ is {\em good} if all values in the set
of values of attributes in $C_2$ for tuples in $S_{ij}$ are modified
to a value in the set in the second update. A set $S_i$ is {\em good}
if it contains a good $S_{ij}$. Sets $S_{ij}$ and
$S_i$ that are not good are {\em bad}. The number of
changes to attributes of $\nit{RHS}(m_2)$
for a bad unblocked $S_i$ is $q(p+q)k + d(p+q)k$, and
for a good unblocked $S_i$ it is at most $q(p+q)k + d(p+q)k - (q+d)$.
Thus the total number of changes for the bad and good
cases is $p^2k + q(p+q)k + d(p+q)k$ and at most
$p^2k + q(p+q)k + d(p+q)k - (q+d)$, respectively.
If the upper bound on the number of changes
from the previous paragraph is taken
as the number of changes for blocked $S_i$,
it is easy to verify that for a given good (bad) $S_i$, the
number of changes when $S_i$ is unblocked (blocked) is strictly
less than the number of changes when $S_i$ is blocked (unblocked).

Consider a sequence $I$ of two updates in which all $S_i$ are
chosen to be unblocked in the first update. Assume that all sets
of values that must be updated to a common value are
updated to a value in the set, except the values of
attributes in $\nit{RHS}(m_2)$ in the first update. We now show
how to improve this pair of updates in order
to obtain a pair of updates leading to an MRI.
For each $j$, there is exactly one $i$ such that
$S_{ij}$ is good. Since all values
of the attributes in $C_2$ occur with the same frequency,
the number of changes resulting from the two updates
does not depend on which $S_{ij}$ are chosen to be good.
The number of changes resulting from applying $I$
to the instance is reduced by changing all bad $S_i$
to blocked. This improvement is maximized by maximizing
the number of bad $S_i$, which can be accomplished by
choosing the set of good $S_i$ so that it corresponds
to a minimum set cover. Denote by $I'$ the pair of updates
obtained by changing $I$ so that it conforms to this
choice of good $S_i$ and by changing all the resulting
bad $S_i$ to blocked.

We now remove the assumption that values from outside
the active domain are used as update values for
attributes in $C_2$ in the first update. This has
no effect on the number of changes for tuples in unblocked
$S_i$, since the first update is ``overwritten" for
these tuples. However, if the update value for a given
equivalence class of $T_{m_2}$ is chosen
as one of the values of a tuple in a blocked $S_i$, it
reduces the number of changes. Let $I''$ be the
sequence of updates obtained by modifying
$I'$ so that each update value for an equivalence class
of $T_{m_2}$ in the first update is chosen from among the
values of tuples in the equivalence class that are in a blocked $S_{i}$.
It is easy to verify that any $I''$ obtained in this
way produces an MRI, and that no other update process will
produce an MRI. Hardness of the pair of MDs now follows from the
fact that the only values that are unchanged in all MRIs
among the values of attributes in $C_2$ are values
in those $S_i$ that correspond to cover sets.
}

\vspace{2mm}
\defproof{Proposition \ref{prop:HSC}}{
We prove the proposition for HSC sets.
In the proof, for an MD $m$, we use the term transitive
closure of $m$, denoted $T_m$, to refer to the transitive
closure of the binary relation that relates
pairs of tuples satisfying the similarity
condition of $m$. For a set of MDs $M$,
the transitive closure of $M$, denoted $T_M$
is the union of the transitive closures of the
MDs in $m$.

Consider an instance $D$ and set of matching
dependencies $M$.
Consider a MD $m$ of the form
\bea
R[\bar A]\approx R[\bar A]\ra R[\bar B]\doteq R[\bar B]\nn
\eea
Let $L$ be the set of all lengths of cycles on the
vertices corresponding to the MDs in $PS(m)$.
Let $n =$ LCM$(L)$ be the {\em period} of $m$.
It is easy
to see that there exists a set $\{S_1,S_2,...S_n\}$ of subsets of $PS(m)$
with transitive closures $\{T_1,T_2,...T_n\}$, where
$\bigcup_i S_i = PS(m)$, such that the following holds.
Let $D_i$ denote an instance obtained by
updating $D$ $i$ times according to $M$,
and for a tuple $t\in D$, denote the
tuple with the same identifier in $D_i$ by
$t^i$. Let $(B,B)$ be a corresponding pair of $(\bar B,\bar B)$.
 After $D$ has been updated $i+a$
times \footnote{We use the term ``update"
even if a resolved instance is obtained
after fewer than $i$ modifications. In this
case, the ``update" is the identity mapping on all
values.}, for $a$ sufficiently large, according to $M$ to
obtain an instance $D_{i+a}$, for all tuples $t$ in
a given equivalence class $E$ of $T_i$,
\bea\label{eq:7}
t^{i+a}[B] = t^{i+a}[B] = v_i^E
\eea
for some value $v_i^E$. Let $D'$ be a resolved
instance. $D'$ satisfies
the property that any number of applications of
the MDs does not change the instance. Therefore,
$D'$ must satisfy (\ref{eq:7}) for all $i$.
That is, for all $1\leq i\leq n$, for any equivalence class $E$ of
$T_i$, and for all tuples $t$ in $E$,
\bea\label{eq:9}
t'[B] = t'[B] = v_i^E
\eea
where $t'$ is the tuple in
$D'$ with the same identifier as $t$.

By (\ref{eq:9}), for any pair of tuples $t_1$ and $t_2$
satisfying \\$T_{PS(m)}(t_1,t_2)$, $t_1'$ and $t_2'$ must
satisfy $T'(t_1',t_2')$, where $T'$ is the transitive
closure of the binary relation on tuples
expressed by $t_1'[B] = t_2'[B]$.
Since the equality relation is closed under
transitive closure, this implies the following property:
\bea\label{eq:11}
T_{PS(m)}(t_1,t_2) \hbox{ implies } t_1'[B] = t_2'[B]
\eea

Equation (\ref{eq:11}) implies that the attribute
values for the tuple/attribute pairs specified in
the proposition must be equal in a resolved instance.
By specifying a series of updates such that only these
values are changed,
we now show that these are the
only changed values in an MRI.

$D$ is updated as follows. For sufficiently large
$a$, after each update attribute $B$ must
satisfy an equation of the form of (\ref{eq:7}) for
each $m$ for which $B\in \nit{RHS}(m)$. Let $T$ be the transitive
closure of the set of all $T_{PS(m)}$ such that $B\in \nit{RHS}(m)$.
For the $(i+a)^{th}$ update,
if the values of $B$ must be modified to enforce
(\ref{eq:7}), use as the common value for all equivalence classes
$E$ contained within a given equivalence class of $T$ the most frequently
occurring value for $B$ in this equivalence class of
$T$. If there is more than one most
frequently occurring value, choose any such value.
After a finite number of updates, an instance is
obtained that satisfies
(\ref{eq:11}).

We must show that
this update process does not change any values other
than those that must be changed to satisfy (\ref{eq:11}).
The theorem will then follow from the fact that
the fewest possible values were changed in order to
enforce (\ref{eq:11}). Let $\{T_1,T_2,...T_{|M|}\}$
denote the set of transitive closures of the MDs
$\{m_1,m_2,...m_{|M|}\}$ in $M$.
For any intermediate instance $I$ obtained in the
update process, let $t_I$ denote the tuple in $I$
with the same identifier as $t$ in the original instance. We will show by
induction on the number of updates that were made
to obtain $I$ that for any $j$, whenever $T_j(t_I,t_I')$
for tuples $t$ and $t'$, it holds that $T(t,t')$. This
implies that updates made to $t[A]$ for any tuple $t$
and attribute $A$ can only set it equal to the common
value for the equivalence class of $T$ to which $t$
belongs.

By definition of $T$, if 0 updates were used to
obtain $I$, $T_j(t_I,t_I')$ implies $T_j(t,t')$ implies
$T(t,t')$. Assume it is true for instances obtained
after at most $k$ updates. Let $I$ be an instance
obtained after $k+1$ updates. Consider the MD
\bea
m_j:~R[A]\approx_j R[A]\ra R[\bar B]\doteq R[\bar B]\nn
\eea
Suppose for the
sake of contradiction that there exist tuples
$t_I$ and $t_I'$ such that
$T_j(t_I,t_I')$ but $\lnot T(t,t')$.
Let $I'$ be the instance of which $I$ is the
updated instance.
Then, there must be a set of tuples
$U = \{t^0,t^1,...t^p\}$ with $t^0 = t$ and
$t^p = t'$ such that $t_I^{i-1}[A]\approx_j t_I^i[A]$
for all $1\leq i\leq p$. By choice of update value, for all $i$,
$T(t^{i-1},s^{i-1})$ and
$T(t^{i},s^{i})$, where $s^{i-1}$
and $s^{i}$ are tuples such that,
$s_{I'}^{i-1}[A] = t_{I}^{i-1}[A]$ and
$s_{I'}^{i}[A] = t_{I}^{i}[A]$. By
$s_{I'}^{i-1}[A]\approx_j s_{I'}^{i}[A]$ and the
induction hypothesis, $T(s^{i-1},s^{i})$.
By transitivity, this implies $T(t^{i-1},t^{i})$
for all $i$, which implies $T(t,t')$, a
contradiction.
}

\vspace{2mm}
\defproof{Theorem \ref{thm:rewrite}}{ We express the query in the
form
\bea\label{eq:form}
\mathcal{Q}(\bar y) = \ex\bar z Q_1(\bar z,\bar y)
\eea
Let $x_{ij}$ denote the variable of $\bar z$
or $\bar y$ which holds the value of the $j^{th}$
attribute in the $i^{th}$ conjunct $R_i$ in $Q_1$.
Denote this attribute by $A_{ij}$.
Note that, since variables and conjuncts can be repeated,
it can happen that $x_{ij}$ is the same variable
as $x_{kl}$ for $(i,j)\neq (k,l)$, that $A_{ij}$ is the same
attribute as $A_{kl}$ for $(i,j)\neq (k,l)$, or
that $R_i$ is the same as $R_j$ for $i\neq j$.
Let $B$ and $F$ denote the set of bound
and free variables in $Q_1$, respectively.
Let $C$ and $U$ denote the variables in $Q_1$ holding
the values of changeable and unchangeable attributes,
respectively.
Let $\mathcal{Q}'(\bar y)$ denote
the rewritten query returned by algorithm {\em Rewrite}, which
we express as
$$
\mathcal{Q}'(\bar y) = \ex z Q_1'(\bar z,\bar y)
$$
We show that, for any constant vector $\bar a$, $\mathcal{Q}'(\bar a)$
is true for an instance $D$ iff $\mathcal{Q}(\bar a)$ is true
for all MRIs of $D$.

Suppose that $\mathcal{Q}'(\bar a)$ is true for an instance $D$.
Then there exists a $\bar b$ such that $Q_1'(\bar b,\bar a)$.
We will refer to this assignment of constants to variables
as $A_{\mathcal{Q}'}$.
From the form of $\mathcal{Q}'$, it is apparent that,
for any fixed $i$, there is a tuple
$t_1 = \bar c_i \equiv (c_{i1},c_{i2},...c_{ip})$ such that
$R_i(\bar c_i)$ is true in $D$ with the following properties.
\begin{enumerate}
\label{enum:one}
\item For all $x_{ij}$ except those in $F\bc C$, $c_{ij}$ is
the value assigned to $x_{ij}$ by $A_{\mathcal{Q}'}$.
\label{enum:two}
\item For all $x_{ij}\in F\bc C$, there is a tuple $t_2$
with attribute $B$ such that $Dup(t_1,A_{ij},t_2,B)$,
and the value of $t_2[B]$ is the value assigned to $x_{ij}$
by $A_{\mathcal{Q}'}$.
Moreover, this value
occurs more frequently than that
of any other tuple/attribute pair in the same equivalence
class of $Dup$.
\end{enumerate}
For any given MRI $D'$, consider the tuple $t_1'$ in $D'$ with the
same identifier as $t_1$. Clearly, this tuple will have the
same values as $t_1$ for all unchangeable attributes, which
by 1., are the values assigned to the variables
$x_{ij}\in U$. Also, by 2.
and Corollary \ref{lem:tc}, for any $j$ such that $x_{ij}\in F\bc C$
is free, the value of the $j^{th}$ attribute of $t_1'$ is
that assigned to $x_{ij}$ by $A_{\mathcal{Q}'}$.

Thus, for each MRI $D'$, there exists an assignment
$A_{\mathcal{Q}}$ of constants to the $x_{ij}$ that makes $\mathcal{Q}$ true, and
this assignment agrees with $A_{\mathcal{Q}'}$ on all $x_{ij}\nin B\bc C$.
This assignment is consistent in the
sense that, if $x_{ij}$ and $x_{kl}$ are the same variable,
they are assigned the same value. Indeed, for $x_{ij}\nin B\bc C$,
consistency follows from the consistency of $A_{\mathcal{Q}'}$, and for
$x_{ij}\in B\bc C$, it follows from the fact that the variable
represented by $x_{ij}$ occurs only once in $Q$, by assumption.
Therefore, $\mathcal{Q}(\bar a)$ is true for all MRIs $D'$, and $\bar a$
is a resolved answer.

Conversely, suppose that a tuple $\bar a$ is a resolved
answer. Then, for any given MRI $D'$ there is a satisfying
assignment $A_\mathcal{Q}$ to the variables in $\mathcal{Q}$ such that
$\bar z$ as defined by (\ref{eq:form}) is assigned the
value $\bar a$. We write $\mathcal{Q}'$ in the form
\bea\label{eq:form2}
\mathcal{Q}'(\bar y)\lar \ex \bar z\w_{1\leq i\leq n}Q_i(\bar v_i)
\eea
with $Q_i$ the rewritten form of the $i^{th}$ conjunct
of $\mathcal{Q}$. For any fixed $i$, let $t' = (c_{i1}',c_{i2}',...c_{ip}')$ be a tuple
in $D'$ such that $c_{ij}'$ is the constant assigned to
$x_{ij}$ by $A_\mathcal{Q}$.

We construct a satisfying assignment $A_{\mathcal{Q}'}$ to the free
and existentially quantified variables of $\mathcal{Q}'$ as
follows. Consider the conjunct $Q_i$ of $\mathcal{Q}'$
as given on line 17 of {\em Rewrite}.
Assign to $\bar v_i'$ the tuple $t$
in $D$ with the same identifier as $t'$. This fixes the
values of all the variables except those $x_{ij} \in F\bc C$,
which are set to $c_{ij}'$. It follows from lemma \ref{lem:tc}
that $A_{\mathcal{Q}'}$ satisfies $\mathcal{Q}'$. Since $A_{\mathcal{Q}}$ and $A_{\mathcal{Q}'}$ match
on all variables that are not local to a single $Q_i$,
$A_{\mathcal{Q}'}$ is consistent. Therefore, $\bar a$ is an
answer for $\mathcal{Q}'$ on $D$.}

\vspace{2mm}

\defproof{Theorem \ref{thm:Chom}}{Hardness follows from the
fact that, for the instance $D$ resulting from the
reduction in the proof of Theorem 3.3 in \cite{Chomicki05}, the set
of all repairs of $D$ with respect to the given key constraint
is the same as the set of MRIs with respect to $m$. The key point is that
attribute modification in this case
generates duplicates which are subsequently eliminated
from the instance, producing the same
result as tuple deletion. Containment is easy.}

\vspace{2mm}
\defproof{Proposition \ref{thm:red}}{Take $\bar A = (A_1,...A_m)$
and $\bar B = (B_1,...,$ $B_n)$. For any tuple
of constants $\bar k$, define
$R^{\bar k}\equiv\s_{\bar A = \bar k}R$.
Let $B_i^{\bar k}$ denote the single attribute
relation with attribute $B_i$ whose tuples are
the most frequently occurring values in
$\pi_{B_i}R^{\bar k}$. That is, $a\in B_i^{\bar k}$
iff $a\in \pi_{B_i}R^{\bar k}$ and there is no
$b\in \pi_{B_i}R^{\bar k}$ such that $b$ occurs
as the value of the $B_i$ attribute in more tuples
of $R^{\bar k}$ than $a$ does. Note that
$B_i^{\bar k}$ can be written as an
expression involving $R$ which is first order
with a $\nit{Count}$ operator.
The reduction produces $(R',t)$ from $(R,t)$,
where
\bea
R'\equiv \bigcup_{\bar k} \l[\pi_{\bar A}R^{\bar k}
\times B_1^{\bar k}\times\cdots B_n^{\bar k}\r]
\eea
The repairs of $R'$ are obtained by keeping,
for each set of tuples with the same key
value, a single tuple with that key value and
discarding all others. By lemma \ref{lem:tc},
in a MRI of $D$, the group $G_{\bar k}$ of tuples such that
$\bar A = \bar k$ for some constant $\bar k$
has a common value for $\bar B$ also,
and the set of possible values for $\bar B$
is the same as that of the tuple with key $\bar k$
in a repair of $D$. Since duplicates are eliminated
from the MRIs, the set of MRIs of $D$ is exactly the
set of repairs of $R'$.}

\vspace{2mm}
\defproof{Theorem \ref{theo:comp}}{$\mathcal{Q}''$ is obtained by composing
$\mathcal{Q}'$ with the transformation $R\ra R'$, which is
a first-order query with aggregation.}

\end{document}